\documentclass{aa}  

\usepackage[switch]{lineno}

\usepackage{graphicx}
\usepackage{txfonts}
\usepackage{natbib}
\usepackage{longtable}
\usepackage{multirow}
\usepackage{dirtytalk}
\usepackage{subcaption}
\usepackage{float}
\usepackage{comment}

\newcommand{\HI}{H\textsc{i}}

\defcitealias{taylor2012_VC1}{T12}
\defcitealias{taylor2013_VC2}{T13}
\defcitealias{dey2025citizen}{D25}

\begin{document}

   \title{Widefield Arecibo Virgo Extragalactic Survey} 

   \subtitle{II. Characterizing the \HI{} properties and environment of the WAVES South region}

   \author{V. Part\'{i}k
          \inst{1,2}
          \and
          R. Taylor
          \inst{2}
          \and
          R. Minchin
          \inst{3}
          }

   \institute{Astronomical Institute of Charles University,
              V Hole\v{s}ovi\v{c}k\'{a}ch 747/2, 180 00 Praha 8, Czech Republic
        \and
             Astronomical Institute of the Czech Academy of Sciences, 
             Bo\v{c}n\'{i} II 1401/1a, 141 00 Praha 4, Czech Republic
        \and
            National Radio Astronomy Observatory, 
            1011 Lopezville Rd., P.O. Box O, Socorro, NM 87801, USA
             }

   \date{Received 7 April 2026 / Accepted 15 July 2026}

  \abstract
   {Galaxy clusters are extreme environments where interactions with the hot intracluster medium drive rapid galaxy evolution. These processes can result in the formation of optically dark gas clouds, as previously observed in Virgo and other clusters.}
   {We investigate the distribution and properties of neutral hydrogen (\HI{}) in two large adjoining regions of the Virgo cluster to understand how the cluster environment influences galaxy transformation. Specifically, we examine the gas content of both star-forming and quiescent populations and search for evidence of gas-loss driven evolution.}
   {We cataloged the 21cm \HI{} Widefield Arecibo Virgo Extragalactic Survey (WAVES) South data using visual and automatic source extraction methods. By combining these results with an optically selected sample, we compared the \HI{} properties of WAVES South with the previously studied VC1 region. To probe gas reservoirs below the nominal detection limit, we performed a stacking analysis of radio spectra across the WAVES South, VC1 and VC2 footprints.}
   {We detected 56 \HI{} sources with a median root mean square (rms) noise of $0.8~\rm{mJy}$, including 50 galaxies, two gas clouds (one being optically dark), and the ALFALFA Virgo 7 complex. Our results reveal a significantly lower detection fraction in WAVES South compared to the VC1 region. Stacking showed no new \HI{} detection at a $0.080~\rm{mJy}$ rms with a maximum of 157 stacked objects from WAVES South, VC1, and VC2.}
   {The lower \HI{} detection fraction suggests that WAVES South is a more dynamically relaxed and evolved environment than the VC1 region. The presence of residual \HI{} in a small subset of early-type galaxies supports a model of dwarf irregular to dwarf elliptical transformation via environmental stripping. Finally, we note a possible evolutionary link between optically dark clouds and recently discovered "blue blobs."}


   \maketitle

    \section{Introduction}
    At an approximate distance of $17~\rm{Mpc}$, the Virgo cluster is the nearest large cluster of galaxies to our own. It contains nearly 2000 galaxies \citep{binggeli1985studies} and is composed of three main infalling subclusters (A, B, and C) and several smaller clouds. The giant elliptical galaxy M87, located in Cluster A, the main body of the cluster, is its most massive member and marks the dynamical center of the cluster. The spatial distribution is complex, as the cluster is likely still assembling its mass, with infalling clouds located as far out as $\sim32~\rm{Mpc}$. Given the diversity and proximity of the cluster, this cluster provides important insights into galaxy evolution, especially when combined with multiwavelength studies.
    
    The member galaxies of the Virgo cluster are, on average, redder than galaxies found in the field, reflecting both their older stellar populations and the likely environmental effects that quench star formation \citep{guiderdoni1985evolution,taylor2012_VC1}. In the core of Virgo, the majority of galaxies are red early-type galaxies (ETGs) \citep{roediger2017next}, with a ratio of roughly two ETGs for every late-type galaxy (LTG), consistent with the morphology–density relation \citep{dressler1980galaxy}. The giant red ETGs generally retain their color with increasing cluster-centric distance, while dwarf ETGs and spiral LTGs tend to become progressively bluer \citep{roediger2011formation}. This is indicative of environmental effects influencing star formation and morphology of the cluster galaxies.

    Given the high relative velocity during a potential galactic encounter in clusters, tidal interactions tend to be less destructive than in the field \citep{taylor2017kinematic}. However, thanks to the presence of intracluster medium (ICM), the ram-pressure stripping (RPS; \citealt{gunn1972infall}) was shown by \citet{vollmer2001ram} to be an effective way of stripping neutral hydrogen (\HI{}) from a galaxy moving through the cluster. The combination of the large number of galaxies in a relatively compact volume, the high angular resolution achievable, and the extreme environment make the Virgo cluster an ideal laboratory for studying these processes.

    The first major optical catalog of the Virgo cluster, the Virgo Cluster Catalog (VCC), was published by \citet{binggeli1985studies}. Covering the entire Virgo cluster and its background, it contains 1776 objects considered as cluster members and is approximately complete to a photographic magnitude of 18. The VCC was later supplemented with the Extended Virgo Cluster Catalog (EVCC; \citealt{kim2014extended}) and the Virgo Cluster Catalogue Additional (VCCA; \citealt{davies2014herschel}), which extended the footprint and the depth of coverage. Modern optical surveys have greatly expanded the coverage and depth of Virgo studies, notably through the Sloan Digital Sky Survey (SDSS; \citealt{kollmeier2025sloan}), the DESI Legacy Survey \citep{dey2019overview}, and the Low-Surface Brightness VCC catalog (LSBVCC; \citealt{davies2015probing}).

    \HI{} observations offer several complementary benefits to optical studies. First, the \HI{} gas, although it does not directly contribute to star formation, it does act as a reservoir for molecular hydrogen and, thus, it is typically found in star-forming galaxies \citep{leroy2008star}. Second, they immediately provide valuable redshift information. Third, low-surface brightness (LSB) galaxies are sometimes much easier to detect in sufficiently deep \HI{} surveys. This is also the case in many peculiar systems both in the Virgo cluster and in field \citep{taylor2016attack,keenan2016arecibo,xu2023discovery}. Furthermore, \HI{} extends further than the stellar component of galaxies and, hence, it provides a useful tracer of environmental effects including interactions with the ICM and with other galaxies. This underscores the importance of \HI{} studies, since they are not optically biased.

    The \HI{} content of Virgo has been studied in several blind 21-cm surveys. Most prominent of these surveys was the Arecibo Legacy Fast ALFA (ALFALFA) survey \citep{giovanelli2005arecibo}. It covered over $7000~\rm{deg^2}$ of the sky, including the entire Virgo cluster. Subsequent targeted surveys traded area for depth, such as the Arecibo Galaxy Environment Survey (AGES), which observed two smaller regions within Virgo at a much greater sensitivity. Because both surveys utilized the same ALFA receiver, they share an angular resolution of $\sim3.5~\rm{arcmin}$ and a raw velocity channel width of $5.5~\rm{km\,s^{-1}}$. At their effective spectral resolution of $10~\rm{km\,s^{-1}}$ (after Hanning smoothing), ALFALFA and AGES achieved average root mean square (rms) noise levels of $2.4~\rm{mJy\,beam^{-1}}$ and $0.6~\rm{mJy\,beam^{-1}}$, respectively.
    
    The first AGES region, known as VC1, was centered on M49 in cluster B, covering a $10\times2~\rm{deg^2}$ area and was described by \citet{taylor2012_VC1} (hereafter, \citetalias{taylor2012_VC1}). A total of 95 cluster members were identified, the majority being late-type galaxies, along with several early-type systems and eight previously unknown \HI{} clouds which \citetalias{taylor2012_VC1} classified as optically dark. The second region, VC2, covers a smaller $5~\rm{deg^2}$ area east of M87. It contains thirteen \HI{} sources, and based on a comparison with VC1, \citet{taylor2013_VC2} (hereafter \citetalias{taylor2013_VC2}) concluded that it is likely still in the process of assembly.

    While it is difficult to determine the typical amount of gas for any individual galaxy, it is known that cluster galaxies are relatively more gas-poor \citep{giovanelli1985gas,solanes2001hi}. The measure of how much gas a galaxy is thought to have lost compared to similar field galaxies is usually described by its \HI{} deficiency \citep{giovanelli1985gas}. Moreover, many galaxies in clusters seem to have no detectable amount of neutral hydrogen. To verify whether this is a physical effect and not just an observational limitation, stacking has been widely used to improve sensitivity. Combining multiple \HI{} spectra in this way has been shown to successfully recover \HI{} below the nominal sensitivity limit in certain regions \citep[e.g.,][]{fabello2011alfalfa,chowdhury2020h,deshev2022arecibo}. In stark contrast, similar efforts failed in the Virgo cluster \citep{taylor2012_VC1,hallenbeck2012gas}, suggesting the nature or intensity of environmental processes is different in Virgo.

    While some objects have clearly lost much of their original gas content, some are strongly or even entirely gas-dominated. As well as the clouds in VC1, \citet{taylor2020faint} discovered a population of optically undetected \HI{} streams indicative of ram pressure. The origins of the clouds are more mysterious. They have typical \HI{} masses of $\sim10^7~\rm{M_{\odot}}$ with velocity widths $W20$ generally around $150~\rm{km\,s^{-1}}$, although some are as narrow as $\sim30~\rm{km\,s^{-1}}$. They are all rather isolated, typically lying more than $100~\rm{kpc}$ from the closest potential parent galaxy, which complicates attempts to explain their origin. \citetalias{taylor2012_VC1} suggested several possibilities, although all with caveats. Together with \citet{taylor2016attack} they favored either the "dark galaxy" hypothesis or the idea that they could be associated with unusually faint stellar systems. They considered tidal interactions to be a possible, but unlikely alternative. Lately, the ram-pressure stripping origin has emerged as the primary suspect, as implied by \citet{jones2024dark}. More recently, \citet{minchin2026high} showed that the dark cloud population is likely not homogeneous and multiple evolutionary scenarios are needed to explain their existence. A related case is the ALFALFA Virgo 7 (AV7) complex, a large \HI{} structure without an optical counterpart, first reported by \citet{kent2007optically} and subsequently investigated by \citet{kent2009arecibo}, \citet{minchin2019widefield}, and \citet{jones2024dark}. 

    Among the 7 ETGs in VC1 detected in \HI{}, \citetalias{taylor2012_VC1} identified three dwarf ellipticals (dEs) that were unexpectedly gas-rich yet showed no structural features. This finding does not fit with the proposed scenario proposed by \citet{boselli2008origin}, whereby dwarf irregulars (dIrrs) are stripped of their gas, most likely through RPS, and subsequently transform into gas-poor dEs. \citetalias{taylor2012_VC1} concluded that these objects have either only recently entered the cluster or represent cases of morphological misclassification. 

    One of the latest discoveries in the Virgo cluster is a potentially new class of stellar systems known as \say{blue blobs} (BBs), first reported by \citet{jones2022young}. These compact systems are extremely low in stellar mass ($\sim10^{5}~\rm{M_{\odot}}$) and dominated by young blue stars with near-Solar metallicities, pointing to a ram-pressure origin. More recently, \citet{dey2025citizen} (hereafter \citetalias{dey2025citizen}) used citizen science to identify 34 additional BBs in Virgo. However, given their novelty and the limited sample size, many questions regarding their nature and origin (for example, their relation to the optically dark clouds and other faint systems in Virgo) remain open. 

    The diversity of unusual systems in Virgo continues to drive the need for deep \HI{} surveys and our limited understanding of the low star-forming systems underscores the importance of the \HI{} studies. One of these important surveys was the Wide Arecibo Extragalactic Virgo Survey (WAVES). Its initial results were reported in \citet{minchin2019widefield}, who concentrated on the analysis of the AV7 system. Here, we present the full results from the WAVES South region of the Virgo cluster, including the complete catalog of the \HI{} detections and their optical photometry. These data do not improve on the study of the AV7 complex and are therefore not discussed in detail in this paper.

    The paper is structured as follows. In Sect.~\ref{sec_obs_data_red}, we describe the observations and the process of data reduction. In Sect.~\ref{section_data_extraction}, we describe the data extraction process and present the object catalogs. We present our results and perform statistical analysis, including stacking, in Sect.~\ref{results}. In Sect.~\ref{summary_discussion}, we summarize and discuss our findings.

\section{Observations and data reduction}\label{sec_obs_data_red}
\begin{figure}
    \centering
    \includegraphics[width=\hsize]{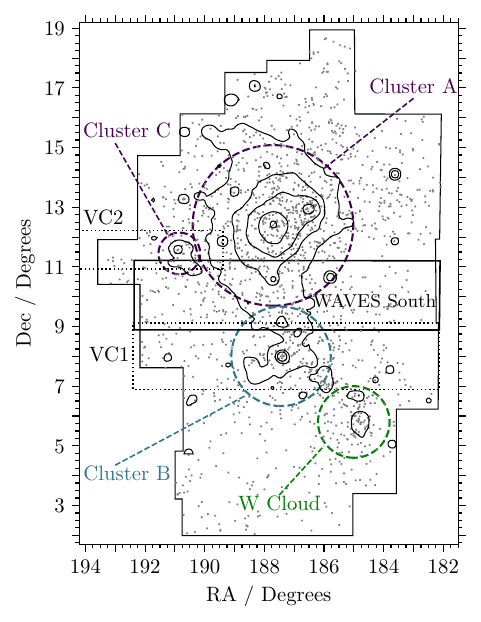}
       \caption{AGES VC1 and VC2 (large and small dotted rectangles, respectively) and WAVES South (solid black rectangle) footprints within the VCC (black outline) in the Virgo cluster. The black contour shows the X-ray gas density from \citet{bohringer1994structure}. Colors of the clusters and clouds represent their assumed distances: purple = 17~Mpc, blue = 23~Mpc, green = 32~Mpc, taken from \citet{boselli2014galex}. Black points indicate VCC cluster members from \citet{binggeli1985studies}.}
          \label{fig_virgo_footprint}
    \end{figure}

    WAVES is a survey conducted with the 305~m Arecibo telescope, which was still ongoing at the time of the telescope's collapse. Its original goal was to build on AGES and survey the remainder of the Virgo cluster to a comparable sensitivity. While the southern region was fully covered, the northern region was only partially observed (around $10~\rm{deg^2}$ out of the planned 30), which will be analyzed in a future paper. In this work, we present the WAVES South data, which ranged from 12:08:26 to 12:49:28 in RA and from +8:52:20 to +11:13:30 in Dec, totaling $20~\rm{deg^2}$. WAVES South centered on the X-ray gas filament between M87 and M49 and is shown along with the AGES VC1 and VC2 regions in Fig.~\ref{fig_virgo_footprint}. On its southern edge, WAVES South overlaps with VC1 by approximately 15~arcmin. 

    As a successor to AGES, WAVES was almost identical in technical implementation and so, the observing setup and data reduction are only briefly summarized here (for further details, see \citealt{auld2006arecibo,minchin2019widefield}). The observations took place in 2017 January and 2018 March and it used the Arecibo L-band Feed Array (ALFA) multibeam receiver in a seven-beam configuration with two polarizations per beam. As with AGES, it utilized the drift scan strategy, where the telescope is fixed in place and the sky is allowed to drift overhead. The beam had a full width at half maximum (FWHM) of 3.5~arcmin. Observations were processed using the Wideband Arecibo Pulsar Processors (WAPP), each with 4096 spectral channels. The total velocity coverage spanned $-2000~\rm{km~s^{-1}}$ to $20\,000~\rm{km~s^{-1}}$ and following a Hanning smoothing, it resulted in a velocity resolution of $10~\rm{km~s^{-1}}$.
     
    The observed data were then processed following the same procedure as in AGES (see \citealt{auld2006arecibo,taylor2012_VC1,taylor2013_VC2}). The data were reduced using the \texttt{AIPS++} packages \texttt{LIVEDATA}, which performed the bandpass estimation and removal, Doppler tracking and calibration of the residual spectrum, and \texttt{GRIDZILLA}, that co-added all the spectra and produced the raw data cube. Finally, a second-order polynomial was fitted to remove baseline ripple, and Hanning smoothing was applied to suppress Gibbs ringing near very bright \HI{} sources \citep{taylor2014arecibo}. Additionally, \citet{minchin2019widefield} discovered that the measured fluxes differed from those previously known - most likely due to a change in receiver gain resulting from damage caused by Hurricane Maria. To account for this, they adjusted the flux measurements using the previously measured sources, resulting in an adjustment of roughly 8\% (for more information see \citealt{minchin2019widefield}). 

    The data quality and interference levels in WAVES South were comparable to those reported in \citetalias{taylor2012_VC1} and \citetalias{taylor2013_VC2}. Several factors contribute to data degradation. The Milky Way provides a significant foreground emission around $\pm 50~\rm{km\,s^{-1}}$, but this lies outside our velocity range (see below) and does not affect the analysis. We also identify a weak, periodic ripple in the baseline, which is likely an instrumental effect caused by imperfectly matched impedance in the telescope wiring following the Hurricane Maria damage. However, it does not significantly affect our results. Aside from the information given above, we do not report any other known man-made radio frequency interference (RFI) in our selected dataset.

    We supplemented the \HI{} observations with optical data from several catalogs. In particular, we made use of the Galaxy On-Line Database Milano network (GOLDMine; \citealt{gavazzi2003introducing}), the Virgo Cluster Catalog, and the NASA/IPAC Extragalactic Database (NED) to achieve the highest possible completeness. For the photometry and radius measurements, we adopted SDSS DR16. The optical counterpart identification was carried out primarily with SDSS and, where necessary, it was complemented by the Legacy Survey DR9. A further discussion of the optical data is provided in the following section.
    
\section{Data extraction and object catalogs}\label{section_data_extraction}
    \subsection{Visual data extraction}
    Detection and extraction of radio sources from data cubes is a complex process and we give only a brief overview here. First, we truncated the velocity range to $300-3000~\rm{km\,s^{-1}}$. This allowed us to avoid the contamination from the Milky Way and High Velocity Clouds and to exclude galaxies at velocities beyond the cluster limit \citep{binggeli1993kinematics}. While this $300~\rm{km\,s^{-1}}$ cut excluded some possible Virgo members with low or blue-shifted velocities (the cluster's velocity distribution extends down to $\sim-700~\rm{km\,s^{-1}}$), available VCC measurements indicate that this omitted range contains less than 13\% of the total cluster population \citep{gavazzi2003introducing}. Next, the flux data cube is converted into S/N, by dividing each spectrum by its own rms value. These cubes are used only for source identification, as the conversion normalizes the rms and makes the cubes easier to inspect and spot sources (see \citealt{taylor2014arecibo}, Fig. A1.). The S/N data cube is then visually inspected using \texttt{FRELLED} \citep{taylor2025frelled}. 

    When a source candidate is found, \texttt{FRELLED} allows the user to interactively mask it, hiding already-cataloged sources, which also records its basic properties, i.e. the coordinates and line width estimates. These parameters are used by the \textit{mbspect} task in \texttt{Miriad} (using the flux cube) to measure systemic velocities, line widths, fluxes and, where possible, to perform position fitting. The flux errors were calculated using the noise rms, signal-to-noise ratio of the peak flux, velocity resolution and a fixed 5\% calibration error, as described in \citet{auld2006arecibo} (Eq. 1), following prescription by \citet{koribalski20041000}.  In cases where the position fitting fails, the coordinates are reverted to the initial by-eye estimates. This occurred for four sources: WCS~20, WCS~36, WCS~38 and WCS~55. We also examined the sample for \HI{} tails and extensions using renzograms, but these results will be published in a future paper. For more information about source extraction we refer to Sect.~2.2 of \citet{taylor2025completeness}.
    
    The \HI{} coordinates of each WAVES Cluster South (WCS) candidate source are then searched in the optical image using the SDSS visual tools. Following AGES established procedure, we search for an optical galaxy within $1.75'$ (half the Arecibo beam) of the \HI{} centroid. When such a galaxy is found, its recession velocity (from NED, if available) is compared with the \HI{} velocity, and if the difference is smaller than $200~\rm{km\,s^{-1}}$ we assign it confidence level~0 (very likely associated with the \HI{} source). A single candidate without a velocity measurement was assigned confidence level~1. Cases with no optical counterpart, or with multiple candidates lacking velocity measurements, were assigned confidence level~2. 
    
    This process provides a consistent way of assigning optical counterparts (OC) to \HI{} detections, which is important both for deriving galaxy properties and for validating the reliability of the \HI{} detection itself. However, while the vast majority of extragalactic \HI{} is associated with optical galaxies \citep{cannon2015alfalfa,kwon2025searching}, some detections do not have counterparts and correspond to genuinely optically dark clouds. Consequently, while we follow this as our standard search criterion, we emphasize that the assignment of an OC is therefore not a strict requirement for inclusion in the catalog, helping us avoid optical bias in our sample.

    AGES was able to make liberal use of follow-up observations since it was a guaranteed part of the awarded observing time. Following the collapse of Arecibo, however, such follow-ups are no longer so readily available, requiring much longer integration times with most other facilities. Therefore, we fall back on other methods to confirm more marginal sources. As well as considering the presence of an optical counterpart, one such is the integrated signal-to-noise criterion \citep{saintonge2007arecibo}. This is defined as 
    \begin{equation}\label{SN_equation}
        {S/N}_{\rm{int}} = \frac{1000F_c}{W_{50}}\frac{w^{1/2}_{smo}}{rms},
    \end{equation}
    where $F_c$ is the total flux in $\rm{Jy\,km\,s^{-1}}$, $W_{50}$ is the line width of the \HI{} source in $\rm{km\,s^{-1}}$, and for $\rm{W_{50}} < 400~km/s$, $w_{smo} = W50/(2\times v_{res})$, where $v_{res}$ is velocity resolution in $\rm{km/s}$. For $W50>400~\rm{km\,s^{-1}}$, $w_{smo} = 400~\rm{km\,s^{-1}}/(2\times v_{res})$, although this was not the case for any of the detected sources in this work. \citet{taylor2025completeness} demonstrated that a S/N threshold of 6.5, as widely used throughout ALFALFA and AGES studies, constitutes a robust threshold for both completeness and reliability. Following the described procedure, in total we cataloged 56 \HI{} sources in WAVES South.

    To verify the completeness of our \HI{} catalog, a second examiner repeated the visual search with previously detected sources masked. In addition to our most confident detections, this second pass revealed twelve additional candidate sources, all of which were marginal (i.e. low $S/N_{\rm{int}}$ and/or lack of OC). Only two of these candidates (WCS~15 and WCS~56) showed a plausible optical counterpart with \HI{} emission centered on them, and were therefore included in the WCS catalog.

    \subsection{Automated data extraction}
    Next, we complemented the two visual searches using the automated SoFiA source finder from \citet{serra2015sofia,westmeier2021sofia} to identify any sources that may have been missed. Based on the findings of \citet{taylor2025completeness}, the following parameters were adopted:
    \begin{verbatim}
        scfind.kernelsXY = 0;
        scfind.kernelsZ = 0, 3, 5, 7, 9, 11, 13,
        15, 17, 19, 21, 23, 25, 27, 29, 31, 33,
        35, 37, 39, 41, 43, 45, 47, 49, 51, 53;
        scfind.threshold = 3.5;
        scfind.replacement = 2.0;
        linker.radiusXY = 2;
        linker.radiusZ = 3;
        linker.minSizeXY = 3;
        linker.minSizeZ = 4.
    \end{verbatim}
    In addition, the reliability parameter was disabled. By trial and error, we iteratively increased the S/N threshold (\texttt{scfind.threshold}) from 3.5 to 3.85 until the algorithm found all our 56 visually cataloged objects, while minimizing the resulting source list. Given that this list is considerably longer than our visual catalogs, this approach is designed to balance the chances of finding plausible new detections while minimizing the number of candidates to inspect.
    
    The extractor identified 211 detections. Subsequently, all remaining sources were evaluated using a similar process as outlined earlier for visual identification. Each spectrum was inspected independently by both observers, who judged its plausibility. Most of these were clearly false detections as per our criteria (primarily extremely weak signals, with low S/N, no OC to support their plausibility and verified by only one observer), but 20 could not be immediately rejected and we are seeking follow-up observations of these objects. They are not included in the analysis.

\subsection{Distance assignments}\label{distance_assign}    
    Given the uncertainties caused by the high velocity dispersion ($\sim650-750~\rm{km\,s^{-1}}$) of the Virgo cluster, the Hubble flow is unreliable for determining distances. Instead, we follow the approach of \citet{gavazzi19993d}, who measured distances of $\sim200$ Virgo galaxies using the Fundamental Plane (FP) and Tully--Fisher (TF) methods. Their method assigns a single representative distance (17, 23, or 32 Mpc) to each subcluster or cloud, which is then applied to all galaxies projected within that structure. These distances are given in the GOLDMine catalog. Following \citetalias{taylor2012_VC1}, we assign distances to our \HI{} detections by comparing with those of their neighboring galaxies and the substructure boundaries. For sources near the boundaries of these structures, velocities were additionally considered when assigning distances. 

\subsection{Catalog of \HI{} detected objects}
    The \HI{} mass of each source was calculated using the standard \HI{} relation,
    \begin{equation}
        M_{\HI{}} = 2.36 \times 10^5 \times d^2 \times F_{\HI{}},
    \end{equation}
    where $d$ is the adopted distance in Mpc and $F_{\HI{}}$ is the integrated \HI{} flux in $\rm{Jy\,km\,s^{-1}}$. 

    The final \HI{} catalog is a combination of two independent visual searches and the SoFiA algorithm, and is shown in Table~\ref{HI_properties_table}. Sources identified by both examiners and SoFiA were retained, while those detected by only one were considered if they satisfied the ${S/N}_{\rm{int}} > 6.5$ criterion. A few sources below this threshold (WCS~15, WCS~49, WCS~51, WCS~52, WCS~55 and WCS~56) are nevertheless included thanks to the presence of a plausible optical counterpart. Thus we regard all of these objects as high-confidence detections.

\subsection{Optical counterparts of WCS galaxies and their photometric parameters} \label{section_optical_counterpart}
    Having identified OCs during our source extraction phase, we now proceed to determine their optical parameters as follows. For each newly detected \HI{} object, we searched the coordinates provided by \texttt{Miriad}'s position fitting (where available) using the SDSS (or, where necessary, Legacy Survey) visual tools. A galaxy located within $1.75~\rm{arcmin}$ radius of the \HI{} position was considered a candidate optical counterpart. When identifying OCs we also accounted for Virgo membership, based primarily on velocity measurements, where available, but also on color and angular size. Cluster \HI{}-bearing objects are typically bluer and exhibit angular radii of $\gtrsim10~\rm{arcsec}$, although this criterion is not absolute. 

    The morphological types (when available) were taken from GOLDMine database (VCC) to ensure a homogeneous classification system. For the nine WCS objects with optical counterparts not listed in GOLDMine, we assigned types manually. For simplicity, we distinguished only four categories: elliptical (0), lenticular (1), spiral (5), and irregular (12). In three cases (i.e., WCS 2, WCS 7, and WCS 36), the \HI{} envelope encompassed more than one galaxy with the same velocity measurement. Visually, these are all pairs of spiral galaxies, but we could not resolve their \HI{} components to determine the properties of each individual object. These systems were excluded from the analysis as we cannot reliably infer the relation between their \HI{} and optical components. A single morphological type could also not be assigned and, thus, the field was left blank. 

    The SDSS photometric parameters were obtained from SDSS DR18 via an SQL query giving apparent magnitudes $g$ and $i$, which are shown in Table~\ref{WCS_optical_table}. The gas clouds WCS~9, WCS 18--21 (AV7) and WCS~54 were excluded as they are not directly associated with any galaxy. For the galaxy size estimates we used the $a$ diameter from GOLDMine, which is defined by \citet{binggeli1985studies} at the faintest visible isophotal surface-brightness level ($\sim25.5~\rm{mag}\,arcsec^{-2}$). For the nine non-VCC galaxies not present in GOLDMine, we fitted an ellipse manually in \texttt{SAOImageDS9}. 

    Identification of the SDSS photometric object actually associated with our \HI{} detection required manual validation. As the query was based only on spatial coordinates it did not always return the most likely optical counterpart. For example, some objects were shown with a radius $<1~\rm{arcsec}$ ($0.08~\rm{kpc}$ at $17~\rm{Mpc}$) compared to the typical $\sim 10~\rm{kpc}$. We identified all such dubious cases and re-ran the query to return all galactic objects within 1 arcmin of the given coordinates. We then manually selected the most plausible counterpart based on the optical size and brightness. In most cases there was a single unambiguous match. When multiple plausible measurements were present, we selected the one with the smallest reported error in magnitude, as these typically corresponded to different measurements of the same object. In a small number of cases, however, the SDSS could not provide a reliable measurement. In those instances we downloaded the SDSS \texttt{.fits} images in the $g$ and $i$ bands and performed independent measurements in \texttt{SAOImageDS9} using the \texttt{funtools} package. This occurred for six \HI{}-detected WCS galaxies and two \HI{} non-detected galaxies.  

    Finally, we subtracted the foreground Galactic extinction values in all photometric bands using the NED Extinction Calculator, which gives the \citet{schlafly2011measuring} recalibration of the \citet{schlegel1998maps} extinction map. The systematic errors are given by the SDSS. The relative photometric uncertainties did not exceed $0.3\%$ in either band. A table with errors is provided separately with this paper.

    \subsection{Catalog of \HI{} non-detected objects}\label{cat_HI_non_det}
    In addition to the \HI{} detections, it was important to construct a comparison sample of optically identified galaxies with no detectable \HI{} emission. This allowed us to investigate differences in gas content, morphology, and environment between \HI{}-rich and \HI{}-poor systems.
    
    To obtain such a sample, we primarily used the GOLDMine catalog. We first selected all galaxies within the WAVES South data footprint with velocities from $300~\rm{km\,s^{-1}}$ to $3000~\rm{km\,s^{-1}}$, consistent with the \HI{} sample. Objects with no velocity measurement were excluded. To ensure the velocity reliability we included only galaxies with high-quality spectroscopy (quality flags 1, 2, and 3; \citealt{gavazzi2003introducing}). Finally, we performed a $1.75~\rm{arcmin}$ cross-match with the coordinates of the optical counterparts of WCS objects, thereby filtering out the \HI{}-detected galaxies. This resulted in a sample of 66 objects without \HI{} detection.

    To increase completeness, we further supplemented this list with a NED query using the same spatial and velocity constraints. In this case, we only included galaxies whose velocities were determined by reliable spectroscopic methods (flags SPEC, SST, SLS and S1L). Objects with UUN or SUN flags (unknown or unknown spectroscopic methods) were excluded, as they correspond to bright and well-known galaxies already present in the GOLDMine catalog. Photometric redshift objects (flag PHOT) were also excluded due to their unreliability. This yielded an additional 113 objects.

    Due to the inhomogeneous nature of NED, some entries were either misclassified as galaxies or contaminated by poor photometry (e.g. foreground star or image artifacts). We visually inspected all 113 objects using SDSS visual tools, discarding unsuitable entries and reducing the number to 102. We then filtered out the galaxies already present in the GOLDMine catalog (VCC, IC and NGC objects) and performed a $1.75~\rm{arcmin}$ cross-match with the WCS catalog, which left 12 galaxies provided by NED.

    The morphological types for the GOLDMine subsample were taken directly from VCC, while for the NED additions we assigned types following the procedure described in the previous section. In total, the catalog of galaxies detected optically but without \HI{} (GOLDMine/NED) in WAVES South contains 78 objects (see Table~\ref{nonHI_properties_table}).

    \section{Results}\label{results}
    In this section, we present our results and compare them to those from the VC1 region. The comparison is motivated by the identical observational setup, the similar sky coverage, and the immediate proximity of the two regions, which share a common boundary (VC2 is too small to provide a meaningful comparison).
    
    \subsection{Statistics of selected galaxy samples}
\begin{figure*}
       \centering
       \includegraphics{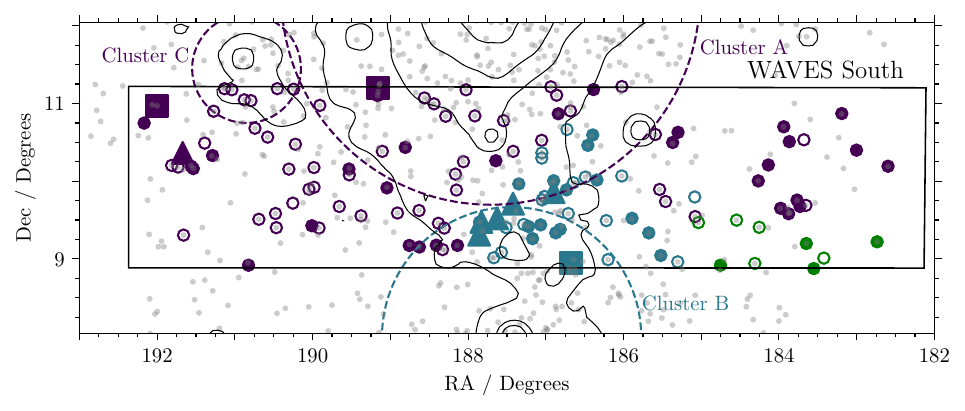}
       \caption{Spatial distribution of \HI{}-detected objects (filled symbols) and optical-only galaxies (open circles). Filled circles indicate galaxies detected in \HI{}, triangles represent gas clouds, and squares mark pairs of galaxies with unresolved \HI{} emission. Symbol color correspond to distance: purple = 17~Mpc, blue = 23~Mpc and green = 32~Mpc. The light gray points show the VCC cluster galaxies from \citet{binggeli1985studies}. The black contours trace the X-ray gas density from \citet{bohringer1994structure}.}
       \label{spatial_distribution_fig}
    \end{figure*}

    In total, we detected 56 WCS objects in \HI{}: 47 individual galaxies (3 early-type and 44 late-type galaxies; see Sect. \ref{sec_morphology}), 3 close pairs of galaxies with unresolvable \HI{} envelopes, the ALFALFA Virgo 7 complex (counted as four sources due to its irregular shape), one gas cloud with ongoing star formation and one optically undetected cloud (we will discuss these below). The derived \HI{} masses range from $8.1\times 10^6~\rm{M_\odot}$ to $1.7\times 10^9~\rm{M_\odot}$. The median rms noise of our measured sources is approximately $0.8~\rm{mJy~beam^{-1}}$ -- $\sim33\%$ higher than the $0.6~\rm{mJy~beam^{-1}}$ reported for AGES \citepalias{taylor2012_VC1}. We attribute this difference to a combination of factors: the partial damage sustained by the Arecibo observatory during Hurricane Maria and the closer proximity of our surveyed area to M87 (a strong continuum source). As discussed in Sect.~\ref{cat_HI_non_det}, we also compiled a sample of 78 optically detected galaxies (66 ETGs and 12 LTGs) with no measurable \HI{} emission (the median rms of the \HI{} non-detected spectra is consistent with the \HI{} detections). The spatial distribution of these WCS and GOLDMine/NED samples, shown in Fig.~\ref{spatial_distribution_fig}, appears relatively uniform. However, the \HI{}-detected galaxies exhibit two noticeable concentrations: one near the center of surveyed data and another toward the western edge between $182 < \mathrm{R.A.} < 185$.

    In addition to our GOLDMine/NED-derived sample of \HI{} non-detections, which are all spectroscopically confirmed cluster members, we also use the VCC to provide a larger comparison sample. The VCC is generally regarded as volume-limited and complete to the 18th photographic magnitude, which is approximately equivalent to the SDSS $g$ band. There are around 300 Virgo-member VCC galaxies located both in the WAVES South (shown in Fig.~\ref{spatial_distribution_fig}) and VC1 footprints. These numbers were obtained by selecting all galaxies from GOLDMine with assigned distances of 17, 23 or 32 Mpc irrespective of their radial velocities, which are not uniformly available across the cluster. This provides a larger, likely more complete sample of Virgo members, but at the expense of reliability. We assume this number to be a reasonable approximation of the true total number of galaxies present.

    As shown in VC1 \citepalias{taylor2012_VC1}, \HI{} surveys have the potential to detect objects missed in optical surveys due to their low surface brightness. Within the WAVES South region, we identified nine galaxies in \HI{} ($\sim16\%$ of the sample) and seven galaxies through the NED query that are not listed in the VCC. Of the nine \HI{}-detected galaxies, seven were later included in the supplementary VCCA catalog \citep{davies2014herschel}, but the remaining two (WCS~10 and WCS~23) remain absent from dedicated Virgo optical surveys. Both systems are low-surface-brightness objects with apparent $g$-band magnitude fainter than the VCC limit, explaining their absence from optical catalogs. Their optical counterparts also lack published spectroscopic redshifts.

    For comparison, VC1 has an \HI{} non-VCC detection fraction of 22\% and VC2 of 39\% (although this is likely due to the low-number statistics with only 13 cluster members). \citetalias{taylor2012_VC1} and \cite{taylor2013_VC2} used only GOLDMine with no NED addition and, hence, their \HI{} non-detected samples were purely VCC galaxies.
    
    \subsection{Comparison with ALFALFA}
    While ALFALFA was not as deep as AGES or WAVES, it is the only Arecibo \HI{} survey to cover the entire Virgo cluster. Within the WAVES South region, we recovered all 48 ALFALFA sources and identified an additional eight \HI{} detections not previously cataloged by ALFALFA \citep{haynes2018arecibo}. These new detections tend to have a lower \HI{} mass, typically $\sim~10^7~\rm{M_\odot}$. The strongest (in terms of $S/N_{\rm{int}}$) source, WCS~10, has an integrated $S/N_{\rm{int}}$ of 11.5. Considering the ratio of the median rms noise between WAVES South ($0.8~\rm{mJy}$) and ALFALFA ($2.4~\rm{mJy}$), we estimate that WCS~10 would have equivalent $S/N_{\rm{int}}$ of only 4.3 in the ALFALFA data - well below the 6.5 detection threshold. The other sources not found in ALFALFA would have even lower S/N levels.

    In VC1, \citetalias{taylor2012_VC1} reported 93 (out of the total 95) Virgo \HI{} detections from AGES data but only 64 ALFALFA cluster members in the $300-3000~\rm{km\,s^{-1}}$ velocity range. Thus, the roughly four times more sensitive AGES survey detected 45\% more sources than ALFALFA. In contrast, in WAVES South the rms noise improved by a factor of about three compared to ALFALFA, yet the number of \HI{} detections increased by only $\sim17\%$. This raises the question as to why the detection rate apparently does not scale directly with sensitivity. One explanation is that our source extraction techniques are missing a large fraction of sources. This is possible but unlikely for two reasons. Firstly, we have searched WAVES South and VC1 in essentially the same way, albeit visualized with different software. Secondly, the source detection experiments described in detail in \citet{taylor2025completeness} (Sect.~4.4.2.) demonstrate that our recovery rates are reliable and consistent. A more likely scenario is that there are simply no faint sources below a certain threshold, and the discrepancy reflects intrinsic differences in the galaxy population and/or local environmental effects within the surveyed volumes. We consider this further in Sect.~\ref{section_stacking}.

    \subsection{\HI{} mass and velocity width distributions}\label{subsec_himass_veldist}
\begin{figure}[h!]
        \centering
            \includegraphics[width=\hsize]{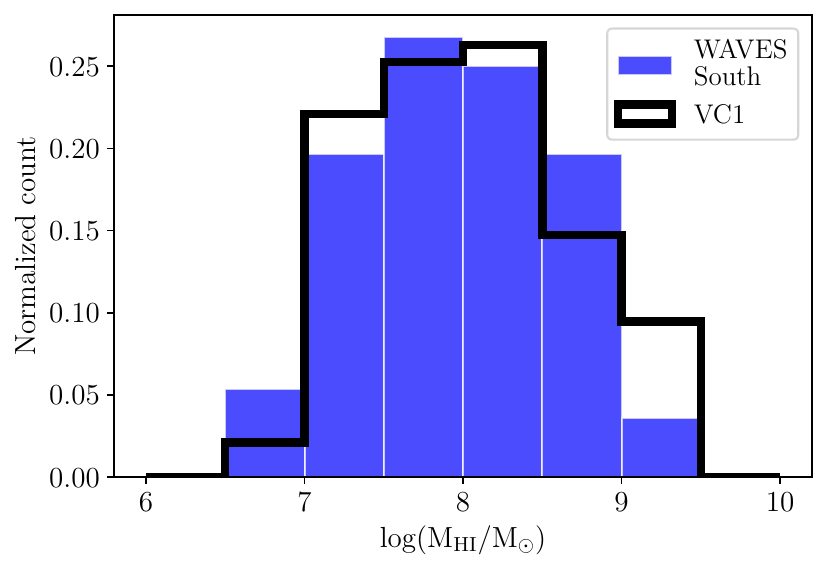}
            \includegraphics[width=\hsize]{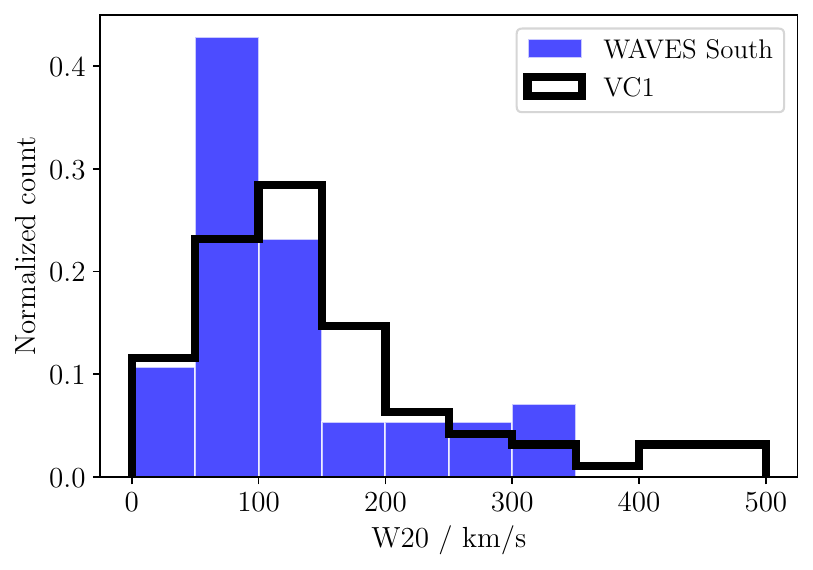}
            \caption{Distributions of \HI{} mass (top) and velocity width at 20\% of the peak signal (bottom) for WAVES South (blue bars) and VC1 (black outline) populations.} \label{mass_width_fig}
   \end{figure}

    The distributions of \HI{} mass and velocity widths for WAVES South and VC1 are shown in Fig.~\ref{mass_width_fig}. The \HI{} mass distributions are nearly identical between the two regions. This indicates that the $\sim33\%$ higher rms noise in WAVES South does not significantly reduce our sensitivity to low-mass galaxies. 
    
    In contrast, WAVES South shows a noticeable excess of galaxies with narrow velocity widths ($W20$ between 50 and $100~\mathrm{km\,s^{-1}}$) but lacks high-width detections (above $350~\mathrm{km\,s^{-1}})$. The median velocity widths are $94.5~\mathrm{km\,s^{-1}}$ for WAVES South and $124.0~\mathrm{km\,s^{-1}}$ for VC1. Statistical tests indicate that this difference is marginal, with the Mann–Whitney (MW) and Kolmogorov–Smirnov (KS) tests yielding $p$-values of 0.07 and 0.06, respectively. Thus, while there is a hint that the WAVES galaxies may have narrower velocity profiles, the evidence is not statistically conclusive.

    \subsection{Velocity distribution}
\begin{figure}[h!]
       \centering
       \includegraphics[width=\hsize]{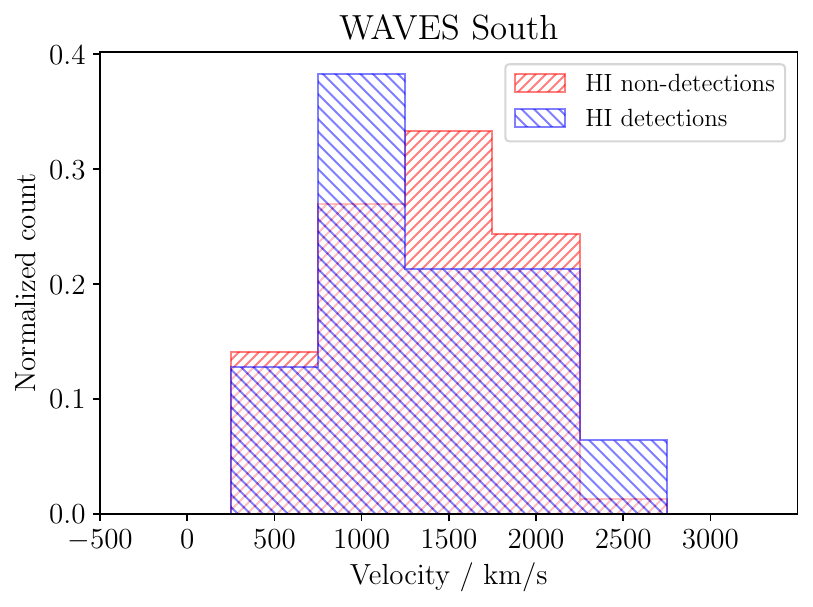}             
       \includegraphics[width=\hsize]{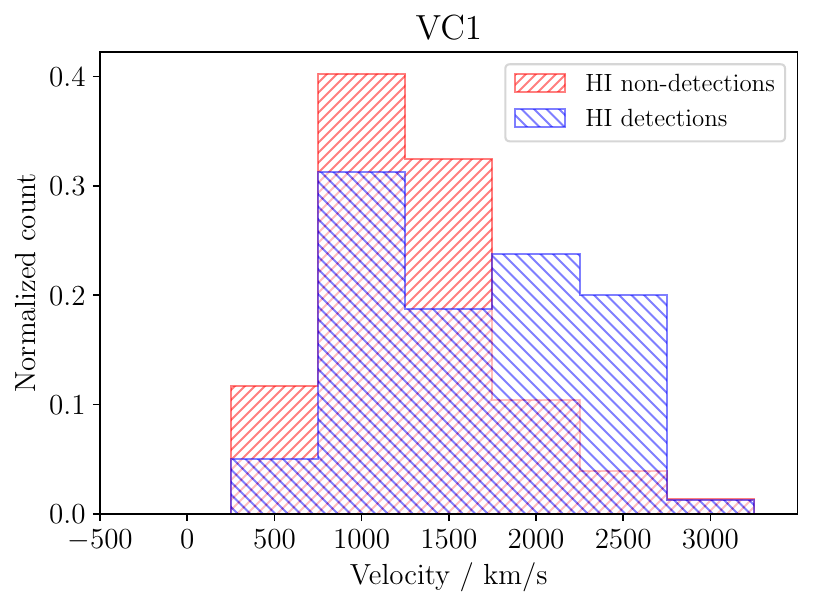}
       \caption{Velocity distributions of the 47 \HI{}-detected galaxies (blue bars) and 78 \HI{} non-detected galaxies (red bars) in WAVES South (top), and 80 \HI{}-detected galaxies (blue bars) and 78 \HI{} non-detected galaxies (red bars) in VC1 (bottom). The bin size is equal to $500~\rm{km\,s^{-1}}$.} 
        \label{vel_distr}
   \end{figure}

    The velocity distributions of \HI{}-detected and non-detected galaxies are shown in Fig.~\ref{vel_distr}. In WAVES South, both populations exhibit a similar shape, velocity range, and mean velocity (around $1350~\rm{km\,s^{-1}}$). This similarity is confirmed by both the MW and KS tests. 
    
    For comparison, we also show the velocity distributions from \citetalias{taylor2012_VC1}, limiting the velocity range to $300-3000\rm{km\,s^{-1}}$ to match the range adopted in this study. In contrast to WAVES South, the VC1 populations are clearly distinct. The \HI{} detections show an approximately uniform distribution, while the non-detections follow a more Gaussian-like profile. This difference is statistically significant, with the MW and KS tests confirming the distinction at the $99.9\%$ and $99.8\%$ significance levels, respectively. As noted by \citetalias{taylor2012_VC1}, the large portion of the VC1 \HI{} population is made of relatively more infalling galaxies or galaxies at higher distances.

    \subsection{Morphology and the color-magnitude diagram}\label{sec_morphology}
    We followed the GOLDMine morphological classification system and considered types from -3 to +2 to be early-type galaxies, with higher values corresponding to late-type galaxies. Type~20 denotes generally unclassified objects; only two such galaxies are detected in \HI{} (WCS~14 and WCS~31). Based on its blue color and the clear structural features visible in the SDSS imaging, WCS~31 (VCC31) is almost certainly an irregular galaxy. It is unclear why it was not classified as such in the GOLDMine database. 
    
    WCS~14 (VCC1295), on the other hand, is more difficult to categorize. It is among the faintest galaxies in our sample, and its reported color ($g-i=0.83$) measured in the SDSS places it near the faint end of the red sequence in the color-magnitude diagram (CMD) in Fig.~\ref{CMD_fig}. However, color measurements for such LSB systems can be susceptible to higher uncertainties in SDSS than in deeper surveys. In fact, as we discuss in Sect.~\ref{sec_res_BB_pop}, \citetalias{dey2025citizen} reports $g-i=0.38$ using the deeper Next Generation Virgo Survey (NGVS) data. While SDSS imaging lacks clear morphological structures for this object, in the Legacy Survey the galaxy appears somewhat irregular. Combined with the UV emission seen in GALEX data and the detection of \HI{}, we conclude that WCS~14 is more likely a late-type dwarf galaxy.
    
\begin{figure}
        \centering
            \includegraphics[width=\hsize]{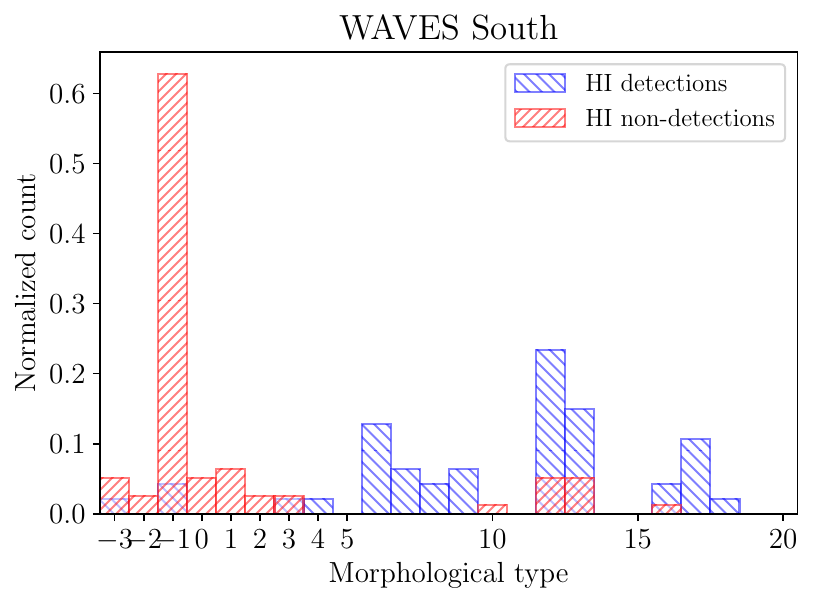}
            \includegraphics[width=\hsize]{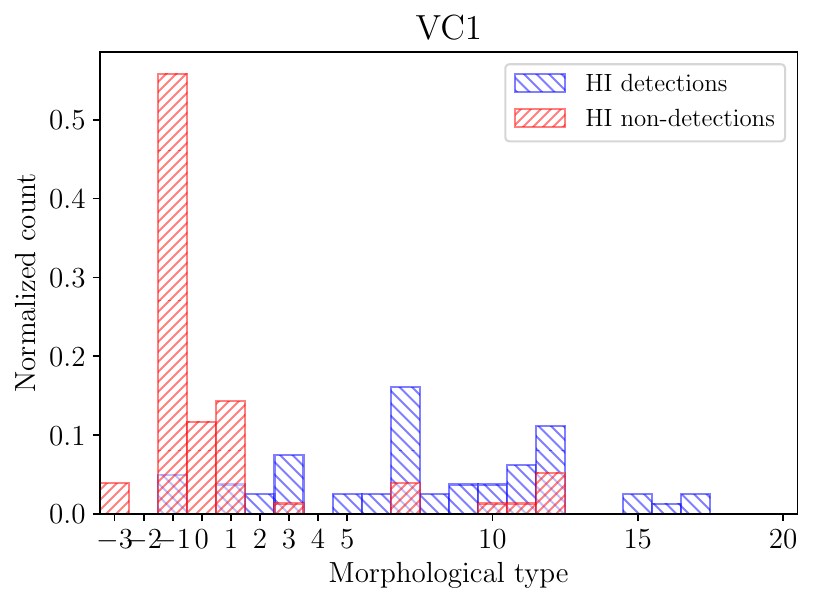}
            \caption{Morphological distributions of \HI{}-detected (blue bars) and non-detected (red bars) galaxies in WAVES South (top) and VC1 (bottom).} \label{morpho_dist_fig}
   \end{figure}

\begin{figure}
        \centering
            \includegraphics[width=\hsize]{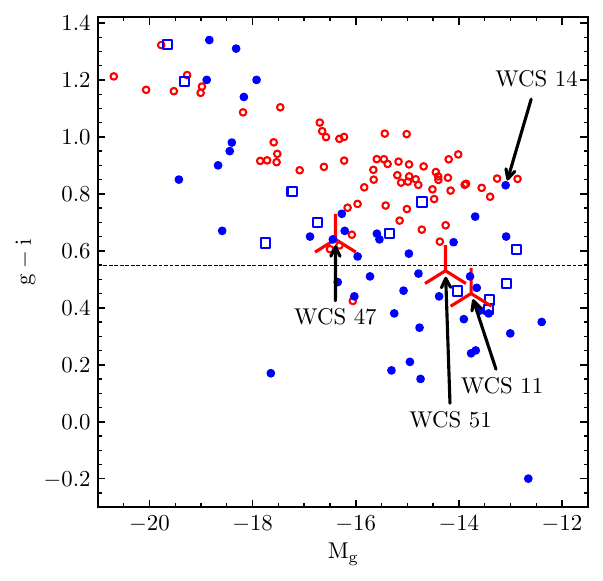}
            \caption{CMD of Virgo-member galaxies in the WAVES South region. Open red circles represent early-type galaxies without \HI{}, open blue squares represent late-type galaxies without \HI{}, red triaxial symbols denote ETGs with \HI{} and blue filled circles denote LTGs with \HI{}. The dashed horizontal line denotes the $g-i=0.55$ level.} \label{CMD_fig}
    \end{figure}

    The morphological distributions of \HI{}-detected and non-detected galaxies for WAVES South and VC1 are shown in Fig.~\ref{morpho_dist_fig}. As expected, the \HI{} detections are dominated by the late-type galaxies, while non-detections are primarily early-type systems. There are, however, several outliers that do not follow this general trend, i.e., LTGs without \HI{} and ETGs with \HI{} detections.

    The 12 late-type galaxies without detectable \HI{} vary significantly in both their physical properties and visual morphology. They can, however, be broadly divided into three distinct groups in the CMD of Fig.~\ref{CMD_fig} (shown as open blue squares). 

    The first group consists of two massive systems, VCC~1412 and VCC~1813, which present clear cases of morphological misclassification (originally classified as spirals). Both are large, luminous ($M_g\approx-19.5$), and red galaxies that exhibit the typical appearance of lenticular systems.

    The second group comprises VCC galaxies that appear generally redder and more passive than the typical \HI{}-bearing LTG population. While their luminosities span several orders of magnitude, they are characterized by $g-i$ colors consistently between 0.55 and 0.85. Despite their redder appearance, several of these objects retain subtle structural features. For instance, although VCC~1273 exhibits a generally smooth disk-like structure, reminiscent of lenticular systems, it also shows possible dust lanes, while VCC~275 displays features similar to those found in elliptical shell galaxies.

    The third group, consisting of SDSS and WISEA objects, is distinctively different. These galaxies are typically fainter ($M_g\approx-14$), bluer ($g-i<0.55$), and more compact than the second group. While they appear relatively smooth, their extreme compactness makes detailed morphological classification difficult, which likely also explains why they were missed by the original VCC.
    
    At least some of these objects could serve to represent galaxies currently within the "green valley" transition region, where they are genuinely redder than the majority of \HI{} detections \citep{cortese2009evolutionary}. Such systems might have only recently been stripped of their gas, retaining sufficiently young stellar populations to exhibit remnants of their former late-type structures. The stark visual differences between these groups suggest that multiple quenching mechanisms may be at play within the cluster environment. 

    Notably, the SDSS/WISEA group is surprisingly similar to the \HI{} non-detected late-type dwarfs (LTDs) reported by \citet{kleiner2023meerkat} in the Fornax cluster. Both populations share comparable stellar masses (their $0.3-1.6\times 10^7~\rm{M_{\odot}}$ to our $0.9-1.9\times10^7~\rm{M_{\odot}}$), as well as similar luminosities and colors. While some of their objects are more diffuse than ours, others appear visually almost identical (e.g. their FSE5 137 and our WISEA J123955.45+095520.4). As \citet{kleiner2023meerkat} suggest, these galaxies could have lost their gas very recently, allowing them to retain their color and structure. This finding stands in contrast to \citetalias{taylor2012_VC1}, where no clear LTGs were identified to be completely devoid of \HI{}, although this was based on an earlier generation of optical surveys.

\begin{figure}[h]
    \centering
        \begin{subfigure}{0.49\columnwidth}
            \centering
            \includegraphics[width=\textwidth]{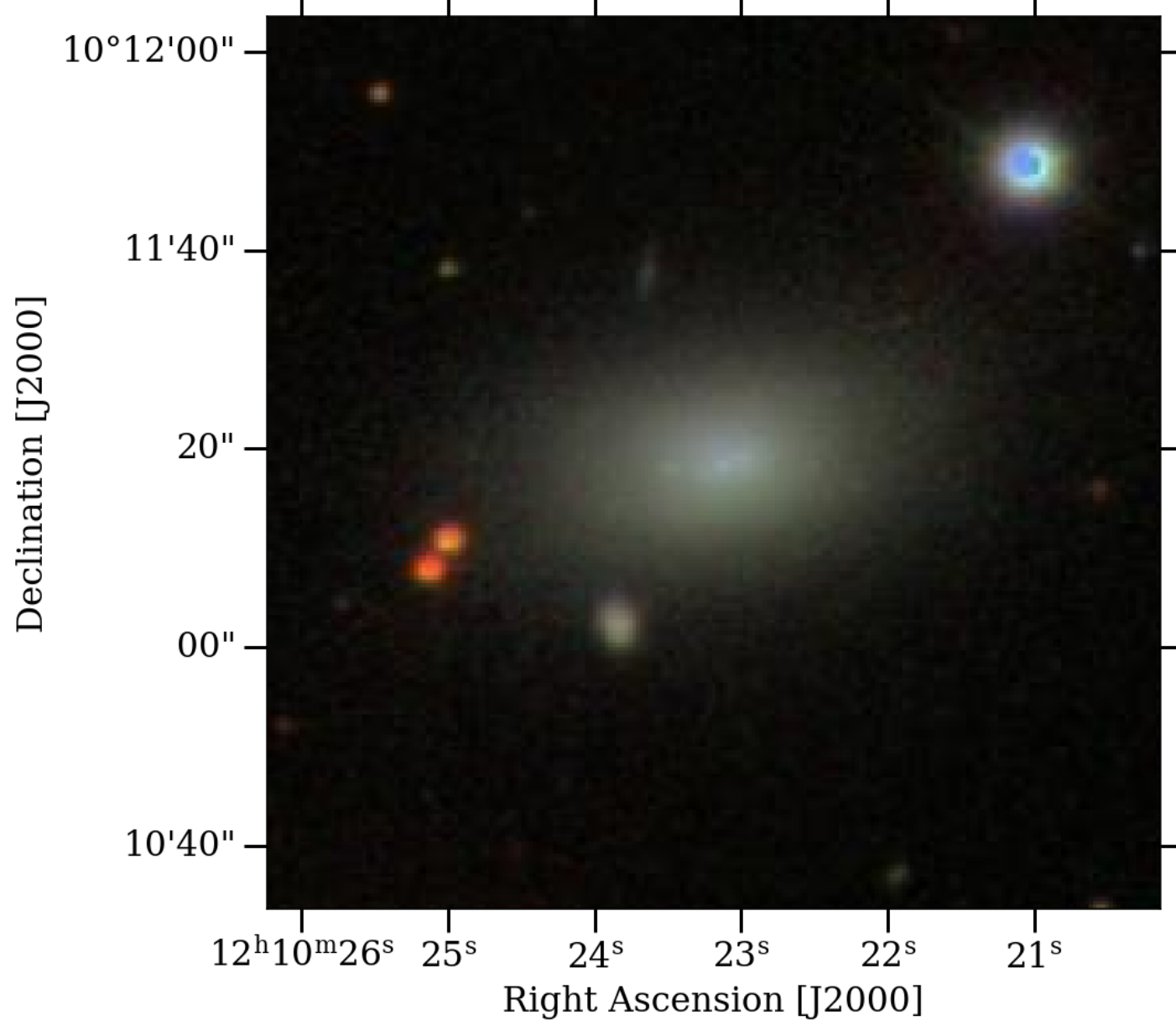}
            \label{fig:wcs47}
        \end{subfigure}
    \hspace{0pt} 
        \begin{subfigure}{0.49\columnwidth}
            \centering
            \includegraphics[width=\textwidth]{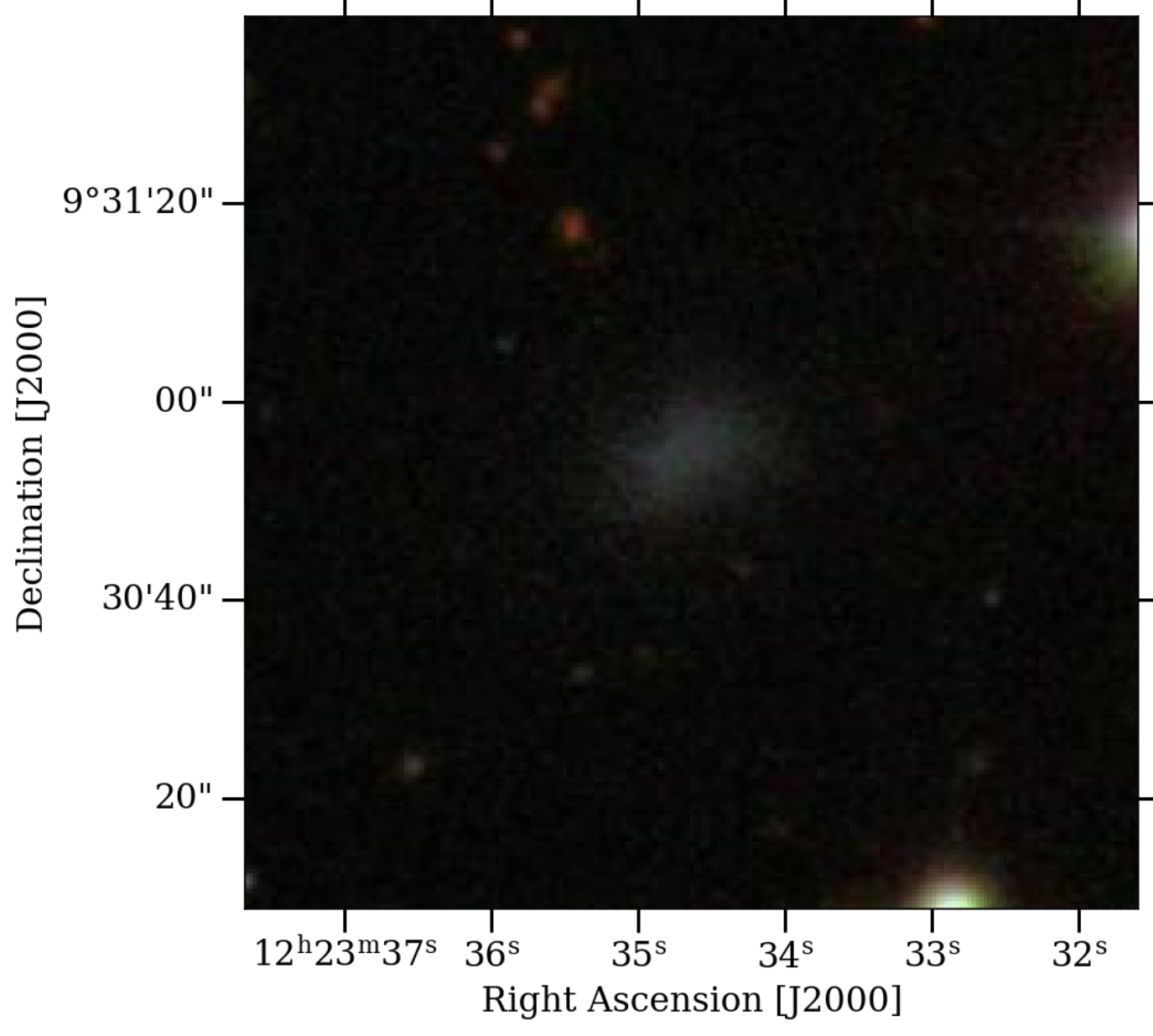}
            \label{fig:wcs51}
        \end{subfigure}
    \caption{SDSS RGB images of early-type galaxies with \HI{} WCS~47 (left) and WCS~51 (right). The field of view is 1.5 arcmin.}
    \label{WCS47_WCS51}
    \end{figure}

    Conversely, we also detected three early-type galaxies in \HI{}: WCS~11 (VCC~1964), WCS~47 (VCC~21), and WCS~51 (VCC~651). These are highlighted in the optical CMD in Fig.~\ref{CMD_fig} as red tri-axial symbols. WCS 47 and 51 are both generally similar in terms of their smooth morphology (shown in Fig.~\ref{WCS47_WCS51}), although WCS~47 has a much brighter central region both in SDSS and Legacy Survey. This seems consistent with their classification as ETGs. Their \HI{} masses are $4.0\times10^7~\rm{M_{\odot}}$ and $1.7\times10^7~\rm{M_{\odot}}$, corresponding to gas fractions of $0.16$ and $0.62$, respectively (for the calculation of stellar masses and gas fractions, see Sect.~\ref{gasfrac_sec}).

    In contrast, WCS~11 appears to be a different class of object altogether and it is likely associated with the ultra-diffuse galaxy candidate VCC~1964 \citep{zaritsky2023systematically}. It also shows a gas fraction that is an order of magnitude higher than the other detected ETGs, though the gas is displaced from the optical center. We discuss this in detail in \citet{taylor2026ultra}, and here only note that the nature of this object is likely quite different to the other ETGs. 
    
    The other galaxies follow the expected bimodal distribution with a narrower red sequence dominated by ETGs and a more dispersed blue cloud of LTGs. All outliers, such as the extremely blue ($g-i\approx-0.2$) WCS~10 or bright-and-blue ($M_g\approx-18, g-i\approx0.2)$ WCS~42, were manually inspected, and their photometric measurements were verified.

    \subsection{Gas fractions}\label{gasfrac_sec}
\begin{figure}
        \centering
            \includegraphics[width=\hsize]{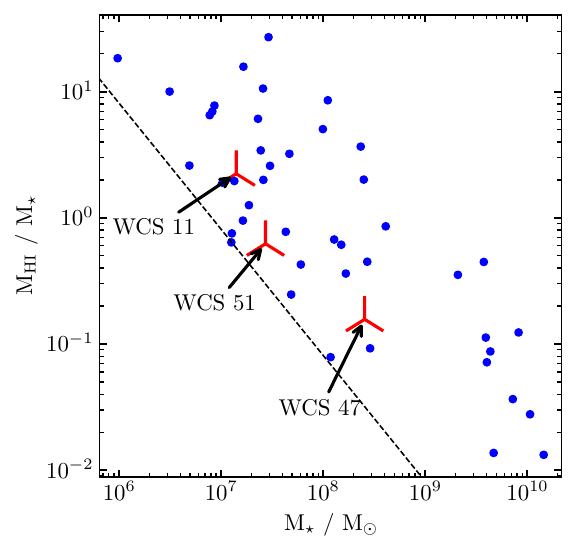}
            \caption{\HI{}-to-stellar mass ratio diagram for the \HI{}-detected galaxies. Blue points represent LTGs, while red triaxial symbols represent ETGs. The dashed line indicates the approximate sensitivity limit, assuming a tophat profile with a median rms of 0.8~mJy, a $3\sigma$ threshold, and a $50~\rm{km\,s^{-1}}$ velocity width at 17~Mpc. Both axes are logarithmic.} 
        \label{gasfrac_fig}   
    \end{figure}

    The \HI{}-to-stellar mass ratio diagram for the \HI{}-detected galaxies is shown in Fig.~\ref{gasfrac_fig}. The stellar mass was calculated using the prescription of \citet{taylor2011galaxy} -- the $i$-band luminosity was calculated using the Pogson equation with an absolute solar magnitude of $M_{\odot_{i}} = 4.58$, and the stellar mass was then derived from
    \begin{equation}
        M_\star = 10 ^{-0.68+0.7\times (g-i)} \times L_i,
    \end{equation}
    where $L_i$ is the $i$-band luminosity.
    
    As found in previous studies, the gas fraction increases at lower stellar masses. This is partially a sensitivity effect: the fainter the galaxy is, the higher the gas fraction needed to detect it. The three \HI{}-detected ETGs exhibit comparable \HI{} masses ($\sim3\times10^7~\rm{M_{\odot}}$), while their stellar masses roughly span an order of magnitude ($\sim10^7-10^8~\rm{M_{\odot}}$). As a result, their gas fractions are systematically lower than those of the LTG population. This is in contrast to VC1, where the \HI{}-detected dEs had typically higher gas fractions. We discuss the implications of this in Sect.~\ref{gas_loss_discussion}.
    
    \subsection{\HI{} deficiency}
    \HI{} deficiency quantifies how much gas a galaxy is likely to have lost compared to a similar field galaxy, which is primarily based on its physical size in kpc. \HI{} deficiency is known to correlate with environment, particularly within clusters, where ram-pressure stripping can effectively remove gas from galaxies. However, it does not indicate whether a gas loss is ongoing or occurred in the past. It is expressed as 

    \begin{equation}
        DEF_{\rm{\HI{}}} = \log M_{\rm{\HI{}_{ref}}} - \log M_{\rm{\HI{}_{obs}}},
    \end{equation}

    where $M_{\rm{\HI{}_{ref}}}$ and $M_{\rm{\HI{}_{obs}}}$ are the predicted and observed \HI{} masses, respectively \citep{giovanelli1985gas}. The predicted mass is that corresponding to a comparable field galaxy in which no environmental gas loss would have occurred, which is based on observations of isolated field galaxies. This is calculated as

    \begin{equation}
        \log M_{\rm{\HI{}_{ref}}} = a + b\times \log d,
    \end{equation}

    where $d$ is the optical diameter of the galaxy in kpc and $a$ and $b$ are parameters based on the typical spiral field galaxies. Several authors report slightly different values, but to maintain consistency with \citetalias{taylor2012_VC1}, we adopt $a=7.51$ and $b=1.460$ from \citet{solanes1995hi}. Given the intrinsic scatter of approximately 0.3~dex, we interpret the results as follows: galaxies with $DEF_{\HI{}} < -0.3$ are gas-rich (negatively deficient), those with $-0.3 < DEF_{\HI{}} < 0.3$ have typical gas content, and those with $DEF_{\HI{}} > 0.3$ are \HI{}-deficient (or strongly \HI{}-deficient if $DEF_{\HI{}} > 0.6$). 

\begin{figure}
        \centering
            \includegraphics[width=\hsize]{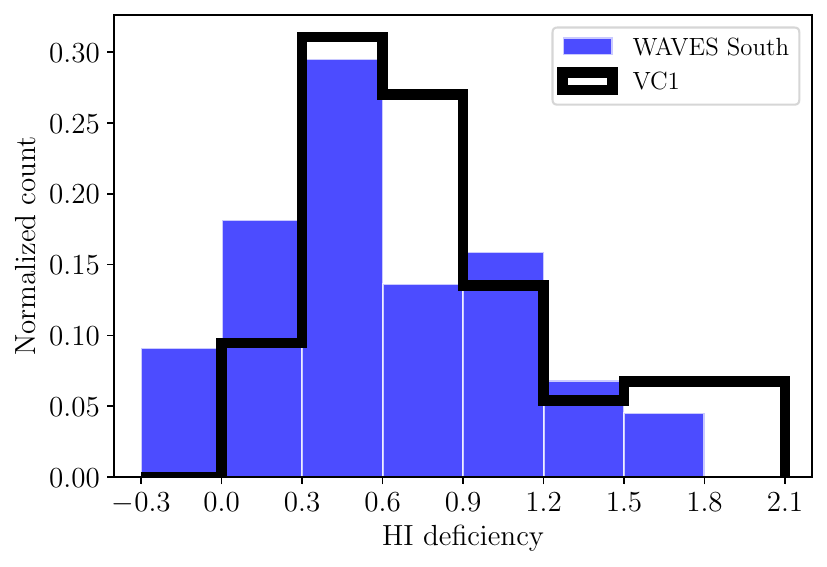}
            \caption{\HI{} deficiency distribution of 44 WAVES South late-type (blue bars) and 74 VC1 (black outline) \HI{}-detected galaxies. The bin size is 0.3.} 
        \label{HI_def_histo_fig}
    \end{figure}

    Following \citetalias{taylor2012_VC1}, we calculate the $DEF_{\HI{}}$ only for the \HI{}-detected late-type galaxies, since early-type galaxies typically lack detectable \HI{}. The resulting distribution for the WAVES South region, compared with Virgo-member galaxies from VC1 is shown in Fig.~\ref{HI_def_histo_fig}. The two distributions are broadly similar but offset by roughly 0.2~dex, with median deficiency of 0.48 for WAVES South and 0.70 in VC1. Both the MW and KS statistical tests indicate that the two populations differ at a 99.3\% confidence level.

\begin{figure}
        \centering
            \includegraphics[width=\hsize]{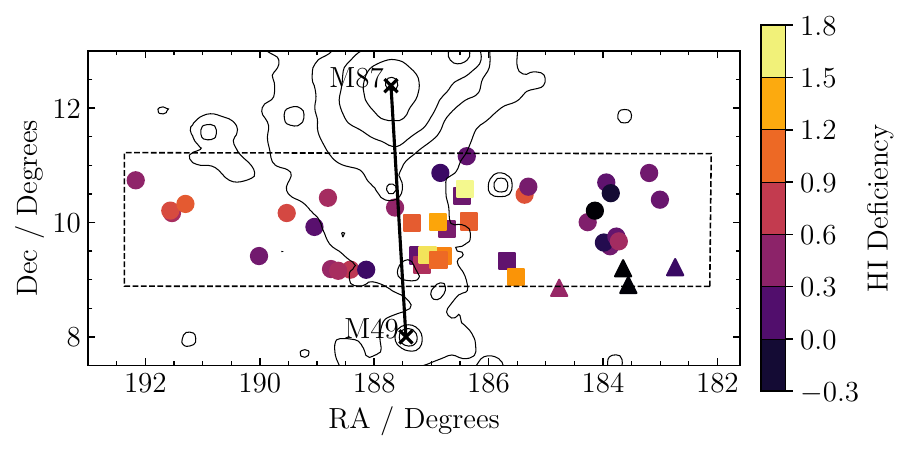}
            \caption{Spatial distribution of \HI{} deficiency in WAVES South. Symbols indicate distance: circles = 17~Mpc, squares = 23~Mpc and triangles = 32~Mpc. The black line marks the projected filament spine of the X-ray gas between M87 and M49.} 
        \label{HI_def_spatial_scatter}     
    \end{figure}

\begin{figure}
        \centering
            \includegraphics[width=\hsize]{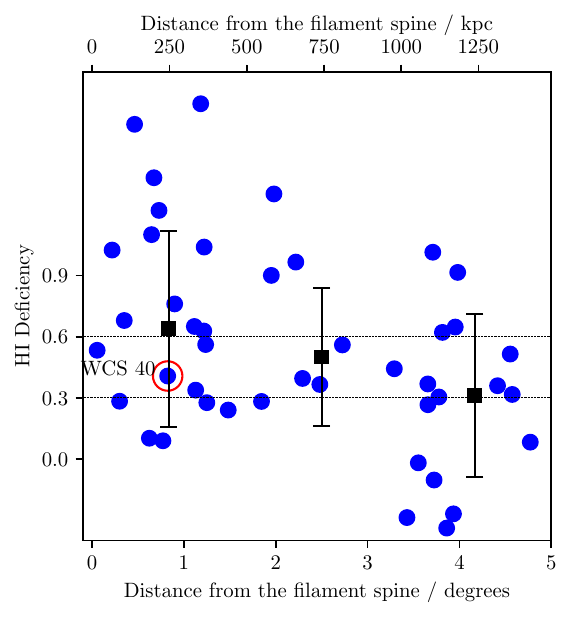}
            \caption{\HI{} deficiency as a function of distance from the line connecting M87 and M49. The projected distance is calculated at the distance of 17~Mpc. The black points show median deficiency in each bin for a better visualization of the declining trend. The error bars were given by the standard deviation of that population. The red circle highlights WCS~40, which is discussed in Sect.\ref{sec_discussion_relax}.} \label{HI_def_dist_fig}    
    \end{figure}

    Since \HI{} deficiency correlates with environment, we plot the spatial distribution of the \HI{}-deficiency levels in Fig.~\ref{HI_def_spatial_scatter}. We also examined the \HI{} deficiency as a function of projected distance from the X-ray filament connecting M87 and M49, assumed to trace the densest part of the intracluster medium. We approximate the filament as a straight line between the two galaxies on the sky (shown in Fig.~\ref{HI_def_spatial_scatter}) and computed the projected separations at an assumed distance of 17~Mpc. The results are shown in Fig.~\ref{HI_def_dist_fig}. 
    
    In another contrast with VC1, where the relation was somewhat bimodal, our plot reveals a steadily declining trend in \HI{} deficiency with increasing distance from the cluster center. The decline is highlighted by the median values in three distance bins and further supported by a Student's t-test, which confirms statistically significant decline at the 99.7\% confidence level. We also note that while choosing a local center is arbitrary to a certain degree, it is plausible that the \HI{} deficiency would correlate with the density of the ICM, which is the highest in the filament spine. This is reinforced by the fact that the most deficient objects are found closer to the denser regions of the ICM and the least deficient are on the cluster outskirts, as seen in Fig.~\ref{HI_def_spatial_scatter}. We also computed the deficiency as a distance from M87, M49, and the data center (average R.A. and declination), which produced similar trends.

    \subsection{Stacking}\label{section_stacking}
    Spectral stacking has been shown to be an effective way to reduce the rms noise and reach an improved sensitivity to constrain the average gas content of the sample without the need for additional observations. To date, various stacking experiments have been conducted across different environments (e.g., the large-scale analysis of approximately 5000 galaxies by \citealt{fabello2011alfalfa}, the more recent study of the Coma cluster by \citealt{healy2021h} or the Abell 1367 field by \citealt{deshev2022arecibo}). However, the application of this method specifically within the Virgo cluster remains relatively limited. Only \citetalias{taylor2012_VC1} (utilizing AGES data) and \citet{hallenbeck2012gas} (utilizing ALFALFA data) have previously targeted this region exclusively. 

    In this procedure, we combined data from WAVES South with the AGES VC1 and VC2 regions, using the same selection method described earlier in this work. Potential cluster members were selected from the GOLDMine catalog using a velocity filter of $300~<v<3000~\rm{km\,s^{-1}}$. To ensure the final stack contained only undetected sources, this list was cross-matched against the \HI{} catalogs of \citetalias{taylor2012_VC1}, \citetalias{taylor2013_VC2} and our current work. We employed a spatial tolerance of 1.75~arcmin and a velocity tolerance of $200~\rm{km\,s^{-1}}$ for the cross-match; any matching sources were excluded from the sample.

    All three data cubes were processed in a consistent manner, which included Hanning smoothing followed by the subtraction of a second-order polynomial baseline. We explored various stacking combinations based on morphology (LTGs, ETGs or a combined sample), region (WAVES South, VC1, VC2), and distance. Regarding the latter, we compared the full sample against a 17~Mpc subset to maximize sensitivity. Each spectrum was visually inspected for both potential \HI{} signals and contamination from Galactic emission. Spectra significantly affected by Milky Way emission or those containing marginal, accidentally included clear detections were discarded. We performed the stacking over both $1000~\rm{km\,s^{-1}}$ and $2000~\rm{km\,s^{-1}}$ windows. The larger window provides a substantial baseline but is more frequently affected by Galactic interference, representing a trade-off between sample size and rms stability.

    We initially performed unweighted stacking, in which spectra are averaged directly. This approach is adequate when all spectra exhibit similar noise levels, as the expected noise in the stacked spectrum scales as
    \begin{equation}
        rms_{stacked} = \frac{rms_{1}}{\sqrt{N}},
    \end{equation}
    where $rms_1$ represents the mean (or median) noise of the individual spectra and $N$ is the number of objects. In practice, however, spectra with significantly higher noise levels can disproportionately degrade the final sensitivity. To mitigate this, we also used variance-defined (also called inverse-variance) weighting. The resulting stacked spectrum $S_{stacked}$ is defined as
    \begin{equation}
        S_{stacked} = \frac{1}{\sum^N_{i=1}w_i}\sum^N_{i=1}S_iw_i, 
    \end{equation}
    where the weights $w_i$ are defined by the noise of each individual spectrum
    \begin{equation}
        w_i = \frac{1}{rms_i^k}.
    \end{equation}
    Following \citet{fabello2011alfalfa}, we adopted $k=2$. Each $rms_i$ value was determined using a $3\sigma$-clipping algorithm before the spectrum was integrated into the stack. To model the expected noise behavior as a function of sample size, we used the following expression,
    \begin{equation}\label{equation_theoretical_curve}
        rms_{theory} = \sqrt{\frac{1}{\sum^N_{i=1}w_i}}.
    \end{equation}
    This allows for a theoretical prediction of the noise based solely on the distribution of individual rms values, independent of the stacking process itself. 

\begin{figure}[t]
        \centering
            \includegraphics[width=\hsize]{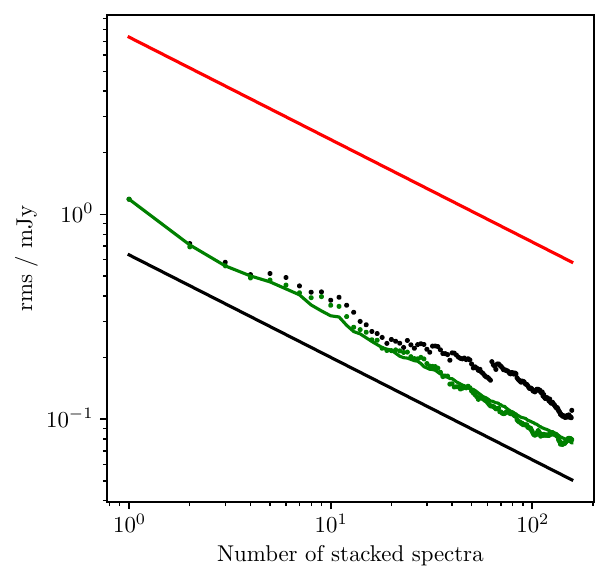}
            \caption{Noise performance as a function of a number of spectra ($N$) for the non-detection sample. The red and black lines represent the theoretical $1/\sqrt{N}$ limits based on the highest and lowest individual rms values in the sample, respectively. The black and green points indicate the measured noise for the unweighted and VD-weighted stacks. The green curve shows the theoretical VD-weighted noise predicted by Eq.~\ref{equation_theoretical_curve}.} \label{stacking_rms_test}
    \end{figure}

    The algorithm proceeds iteratively. Starting with a single spectrum, we compute its clipped and theoretical noise (which are identical at $N=1$). We then add spectra one by one, updating the theoretical and measured rms values and performing a new variance-defined (VD) weighted stack at each step. Comparing the actual measured rms to the prediction from Eq.~\ref{equation_theoretical_curve} serves both as a sanity check and a useful diagnostic of the stack's performance. Fig.~\ref{stacking_rms_test} illustrates this behavior for a representative stack of 157 spectra, encompassing all regions, distances and morphologies.
    
    Fig.~\ref{stacking_rms_test} illustrates several key characteristics of our stacking procedure. The VD-weighted approach consistently outperforms unweighted stacking, yielding lower rms values for an equivalent number of spectra. Furthermore, the theoretical curve (shown in green) closely tracks the VD-weighted results. This confirms that Eq.~\ref{equation_theoretical_curve} provides a reliable performance estimate without the need to construct the full stack. We also explored the normalized weighting schemes (where weights are calculated as $w_i=rms_i/rms_{\rm{min}}$); however, their implementation added significant computational complexity without providing a substantial improvement in sensitivity over the VD-weighted approach.

    While no new \HI{} detections were identified through this stacking analysis, the method allowed us to establish stringent upper limits on the gas content for several subsamples across WAVES South, VC1 and VC2. Additionally, we calculated the \HI{} mass estimates using the \HI{} luminosities, following \citet{fabello2011alfalfa} (Sect.~3.3, Eq.~5), which take into account distances of individual objects (for more see their Sect.~3.3). These results are summarized in Table~\ref{stacking_table}. 

    \begin{table*}[h!]
        \centering
        \small
        \caption{Summary of stacking results.}
        \label{stacking_table}
            \begin{tabular}{clccccc}
            \hline\hline
            Spectral                           & Data                         & Morphology & $N$ & rms     & Distance       & $M_{\HI{}}$                      \\
            width                              &                              &            &     &[$\rm{mJy}$]& [$\rm{Mpc}$]  & [$\times 10^6~\rm{M_\odot}$]     \\
            \hline
        \multirow{10}{*}{$1000~\rm{km\,s^{-1}}$} & WAVES                        & LTG        & 10  & 0.399     & 21.2           & 6.84                             \\
                                               & WAVES                        & ETG        & 58  & 0.166     & 18.9           & 2.25                             \\
                                               & WAVES                        & Both       & 68  & 0.145     & 19.2           & 1.99                             \\
                                               & WAVES, VC1 and VC2           & LTG        & 17  & 0.277     & 23.4           & 7.10                             \\
                                               & WAVES, VC1 and VC2           & ETG        & 139 & 0.091     & 19.3           & 1.34                             \\
                                               & WAVES, VC1 and VC2           & Both       & 157 & 0.080     & 19.8           & 1.26                             \\
                                               & WAVES at 17 Mpc              & LTG        & 6   & 0.537     & 17.0           & 5.49                             \\
                                               & WAVES at 17 Mpc              & ETG        & 43  & 0.170     & 17.0           & 1.74                             \\
                                               & WAVES at 17 Mpc              & Both       & 49  & 0.152     & 17.0           & 1.56                             \\
                                               & WAVES, VC1 and VC2 at 17 Mpc & Both       & 104 & 0.097     & 17.0           & 1.00                             \\
            \hline
        \multirow{10}{*}{$2000~\rm{km\,s^{-1}}$} & WAVES                        & LTG        & 9   & 0.411     & 21.7           & 7.28                             \\
                                               & WAVES                        & ETG        & 45  & 0.160     & 18.6           & 2.26                             \\
                                               & WAVES                        & Both       & 53  & 0.146     & 19.0           & 2.08                             \\
                                               & WAVES, VC1 and VC2           & LTG        & 13  & 0.310     & 24.2           & 8.15                             \\
                                               & WAVES, VC1 and VC2           & ETG        & 84  & 0.103     & 19.7           & 1.71                             \\
                                               & WAVES, VC1 and VC2           & Both       & 97  & 0.100     & 20.3           & 1.73                             \\
                                               & WAVES at 17 Mpc              & LTG        & 5   & 0.541     & 17.0           & 5.53                             \\
                                               & WAVES at 17 Mpc              & ETG        & 36  & 0.173     & 17.0           & 1.77                             \\
                                               & WAVES at 17 Mpc              & Both       & 41  & 0.164     & 17.0           & 1.67                             \\
                                               & WAVES, VC1 and VC2 at 17 Mpc & Both       & 60  & 0.126     & 17.0           & 1.29                             \\
            \hline
            \hline
            \end{tabular}%
    \tablefoot{The samples include objects from all distances unless specified otherwise. $N$ denotes the number of stacked spectra. The distance corresponds to the mean value of all objects in a given subsample. The \HI{} mass sensitivities were derived using the prescription by \citet{fabello2011alfalfa}, and assuming a $3\sigma$ detection threshold and a $50~\rm{km\,s^{-1}}$ top-hat profile.} 
    \end{table*}    

    Our stacked mass sensitivity reaches an order of magnitude below our nominal detection limit. These results reinforce the conclusions of \citetalias{taylor2012_VC1} regarding the efficiency of gas stripping mechanisms in Virgo. The absence of an \HI{} signal, even in our deepest aggregate stacks, suggests that once a galaxy's neutral gas content is reduced below the detection limits of surveys such as ALFALFA or AGES, it is rapidly and effectively removed. 

    \subsection{Other notable objects}
    Apart from the already mentioned ETGs detected in \HI{} we also found several more \HI{} objects that deviate from the general trends and merit individual discussion. These include gas clouds with no apparent host galaxy and a recently discovered class of low-mass stellar systems called BBs \citep{jones2022young}. There are currently 40 known BBs in the \citetalias{dey2025citizen} catalog -- 14 with known \HI{} content. Out of these 14, WAVES South and VC1 contain 10 -- a majority of the \HI{} population.

    \subsubsection{Blue blob population in WAVES South}\label{sec_res_BB_pop}
\begin{figure}[h!]
        \centering 
            \includegraphics[width=0.5\textwidth]{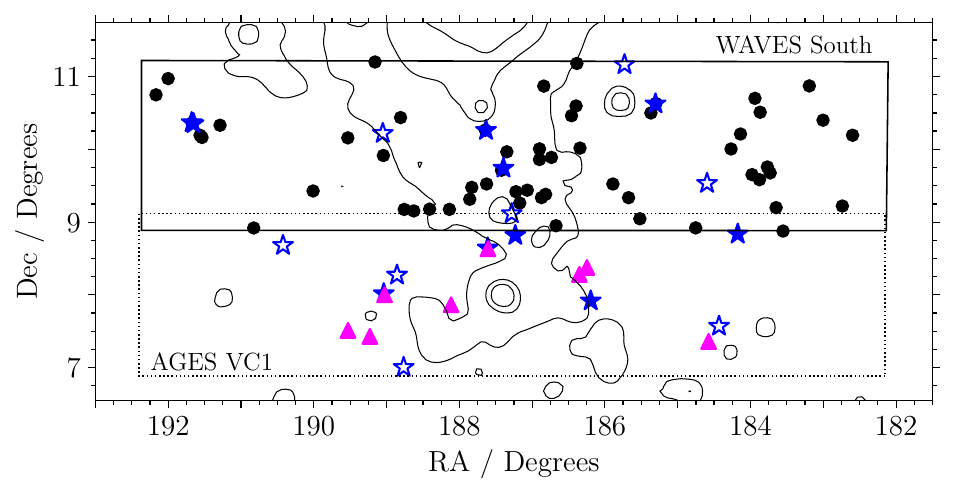}
            \caption{Spatial distribution of BBs from \citetalias{dey2025citizen} in WAVES South and VC1. The filled and unfilled blue stars show the BBs with and without the \HI{}, respectively. The black points show the \HI{} sources detected in this work and the magenta triangles show the VC1 dark clouds from \citetalias{taylor2012_VC1}.}\label{WAVES_VC1_BBs_dist} 
    \end{figure}

    The WAVES South region contains nine BBs from the \citetalias{dey2025citizen} catalog. Five of these are associated with previously known ALFALFA \HI{} sources, all of which are recovered in our WAVES data (the remaining four lack detectable \HI{} emission). Notably, four out of five BBs with \HI{} in WAVES South fall within the uppermost quartile of \HI{} masses for this population. As shown in Fig.~\ref{WAVES_VC1_BBs_dist}, the spatial distribution of these sources shows no clear trend across the sky; however, as noted by \citetalias{dey2025citizen}, on a large scale the BB populations generally avoid both the extreme cluster outskirts and the central region surrounding M87. 

    Our visual inspection of the region revealed no obvious parent galaxy candidates for any of the identified BBs. While some BBs are spatially adjacent to other \HI{} systems, none of the neighboring galaxies exhibit clear signs of recent or ongoing gas removal. However, some BBs are located near the boundaries of our survey area, making a complete environmental inspection impossible. 

    The \HI{} masses reported by \citetalias{dey2025citizen} generally align with our measurements, despite some individual scatter. Our reported mass for WCS~18 (BC6) is three times lower, likely because \citetalias{dey2025citizen} integrated the entire AV7 complex; including the full complex reduces this difference to a marginal level. The most significant discrepancy involves WCS~46 (BC40), which \citetalias{dey2025citizen} report as six times more massive than our estimate. While our total \HI{} flux measurements agree, the distance assigned in the ALFALFA catalog is $40~\rm{Mpc}$, instead of $\sim17~\rm{Mpc}$, used by \citetalias{dey2025citizen} and us. Correcting for this distance scaling brings the \HI{} mass back into good agreement. For other common detections, such as WCS~14 (BC33) and WCS~9 (BC3/BC12), the reported masses agree within a factor of two. 
    
    Of the five \HI{}-detected BBs in WAVES South, three are associated with systems previously identified as isolated gas clouds and two with systems identified as dwarf galaxies. The non-homogeneous nature of the BB population is exemplified by WCS~9 (AGC~226178, e.g. \citealt{jones2022agc}), a unique \HI{} cloud that hosts two separate BBs (BC3 and BC12) rather than a single stellar counterpart. We also note the presence of VCC~1295 (BC33 or WCS~14 in this work), which we identify as a likely dIrr galaxy (see Sect.~\ref{sec_morphology}). Its stellar mass estimates vary depending on the prescription used. Due to limitations of SDSS sensitivity with low-surface brightness objects, we refer the more accurate analysis to \citetalias{dey2025citizen}, where they report a stellar mass of $\sim4\times10^6~\rm{M_{\odot}}$ (our own estimate is approximately four times higher). In either case, it remains the most massive blue blob identified in \citetalias{dey2025citizen} catalog, suggesting that the BB population may extend into the regime of more typical dwarf galaxies. Additionally, given the similar \HI{} masses and other properties, BBs may also be related to the optically dark \HI{} clouds, as discussed below.

    \subsubsection{Verification of previously discovered dark clouds in WAVES South}
    In addition to the ALFALFA Virgo 7 complex, at least two other \HI{} cloud candidates have been reported within the WAVES South footprint prior to our observations. The first is a proposed significant extension to the \HI{} tail of NGC~4424. While the main tail was originally observed by \citet{chung2007virgo}, \citet{sorgho2017h} later reported a much larger extension based on a combination of KAT-7 and Westerbork Synthesis Radio Telescope (WSRT) data, however, this extended feature was only visible in the KAT-7 observations. Although our WAVES data reaches a deeper column density sensitivity than their combined dataset, we do not recover this extension. As previously noted by \citet{minchin2021widefield} (see their Fig.~1), follow-up single-pointing observations with the Arecibo L-Band Wide receiver also failed to detect it. We therefore conclude that this extended tail is likely an artifact of the KAT-7 data rather than a physical feature.

    The second candidate is the KW Cloud, similarly reported by \citet{sorgho2017h} near NGC~4451. Although the reported properties of this source place it well above the detection limits of both WAVES and the earlier ALFALFA data \citep{kent2007optically}, it is not recovered in either survey. Consistent with the non-detections in the targeted Arecibo follow-up observations presented in \citet{minchin2021widefield} (see also their Fig.~1), we find no evidence for the KW Cloud in our dataset, strongly suggesting that this, too, is an artifact in the KAT-7 data.

    \subsubsection{The sole dark cloud candidate}\label{subsec_wcs54}
\begin{figure}
        \centering
            \includegraphics[width=\hsize]{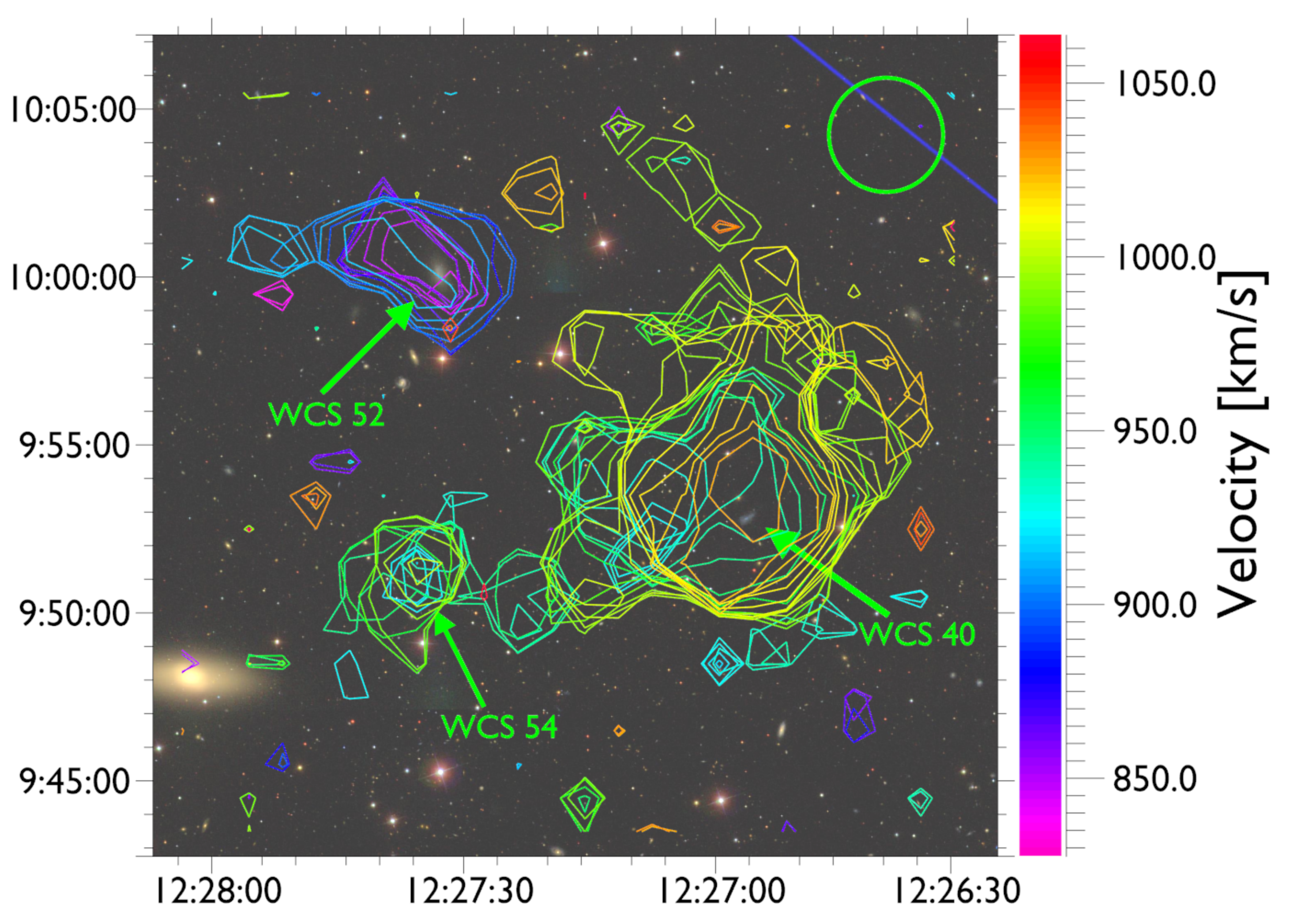}
            \caption{$3\sigma$ renzogram of WCS~54 after Hanning smoothing along the velocity axis. The background is an SDSS RGB image with enhanced brightness and contrast. The \HI{} bridge connecting WCS~54 and WCS~40 is clearly visible. The green circle in the upper right corner is the Arecibo 3.5 arcmin beam.} \label{WCS_54_renzo}    
    \end{figure}

    WCS~54 is the only candidate dark \HI{} cloud identified in the WAVES South region. Located at an assumed distance of 23~Mpc, it exhibits no detectable optical counterpart. WCS~54 has a systemic velocity of $954~\rm{km\,s^{-1}}$, an \HI{} mass of $3.6\times 10^7~\rm{M_{\odot}}$, and a velocity width of $W20=85~\rm{km\,s^{-1}}$. Following the methodology of \citet{taylor2022arecibo}, we present a renzogram of the systems in Fig.~\ref{WCS_54_renzo}. This visualization utilizes Hanning smoothing over five velocity channels to enhance the visibility of structures with sufficient velocity width.

    The renzogram reveals a clear \HI{} bridge connecting WCS~54 to the nearby \HI{}-detected galaxy WCS~40 (VCC~952), situated approximately 10~arcmin (66~kpc in projection) away. Given their spatial proximity, similar recession velocities, and the connecting bridge, these two systems are likely associated and may form a contiguous tidal or ram-pressure tail. However, it should be noted that the bridge is clearly visible only after Hanning smoothing - in the initial inspection WCS~54 resembles the discrete clouds in VC1. 
    
    The gas distribution of WCS~40 appears highly disturbed, characterized by compressed contours to the southwest and a secondary tail-like extension to the north. This morphology is consistent with an ongoing ram-pressure stripping \citep[e.g.,][]{luo2023tracing}. This scenario is further supported by the nearby late-type galaxy WCS~52 (VCC~1021), also shown in Fig.~\ref{WCS_54_renzo}. WCS~52 is located at a similar distance and displays a marginal tail pointing in the same general direction as the WCS~40 extensions. While the emission in WCS~52 is significantly less complex, this may be attributed to its extremely high \HI{} deficiency ($DEF_{\HI{}}=1.38$).

    Although WCS~54 shares similar \HI{} properties with the dark clouds found in VC1, its clear association with a parent galaxy distinguishes it from the VC1 population (typically residing more than 100~kpc from the nearest potential parent galaxy). Tentatively, WCS~54 might represent an early stage in the formation of such systems, potentially providing a link to the BBs or the isolated \citetalias{taylor2012_VC1} clouds. Alternatively, it may simply be a transient, high-density knot within a stripping tail that will disperse into the ICM within a few million years. High-resolution observations are needed to ascertain whether there are ordered motions within WCS~54, as these would imply stability and a longer lifetime.

    \section{Summary and discussion}\label{summary_discussion}
    We utilized 21~cm \HI{} data from the Arecibo telescope to survey a $20~\rm{deg^2}$ region of the Virgo cluster (WAVES South), centered on the X-ray filament between M87 and M49. This survey achieved a median rms of $0.8~\rm{mJy}$, resulting in the detection of 56 \HI{} sources, 47 of which are individual galaxies with clearly identified optical counterparts. For a comparative analysis, we also compiled a sample of 78 optically detected galaxies within the same footprint that lack detectable \HI{}.

    By comparing the WAVES South results with the adjacent VC1 region, which was studied with an identical setup in \citetalias{taylor2012_VC1}, we found significant evidence that WAVES South is a dynamically more relaxed environment containing a more evolved galaxy population. Furthermore, our results provide evidence of gas-loss driven evolution in dIrr galaxies, allowing us to establish robust \HI{} mass constraints on undetected populations via a stacking analysis. We also note a possible connection between the BBs and the VC1 dark clouds.

    \subsection{Dynamical relaxation of WAVES South}\label{sec_discussion_relax}
    Several lines of evidence suggest that WAVES South is more evolved than VC1. First, we compared the velocity distributions of the galaxy populations (Fig.~\ref{vel_distr}). In the VC1 region, \HI{}-detected and non-detected galaxies exhibit clearly distinct velocity profiles. In contrast, both populations in WAVES South show remarkably similar ranges and shapes, suggesting that the gas-rich and gas-poor systems have had sufficient time to reach a state of dynamical mixing. 

    Second, there is a significantly lower fraction of \HI{}-bearing galaxies in WAVES South compared to VC1. The \HI{} detection fraction for VCC galaxies $(\frac{N_{\HI{},VCC}}{N_{VCC}})$ in WAVES South is only $\sim14\%$, compared to $\sim22\%$ in VC1 (both areas contain a similar number of VCC galaxies). When the \HI{} detections associated with all galaxies in general (VCC and others; $\frac{N_{\HI{},VCC}+N_{\HI{},non-VCC}}{N_{VCC}+N_{\HI{},non-VCC}}$) are considered this disparity is even more pronounced ($16\%$ vs $28\%$), confirming that the low detection rate is a physical effect rather than a statistical artifact. As discussed previously, we assumed the number of VCC galaxies is a reasonable approximation of the true total number of galaxies. More details are given in Sect.~4.1 of \citetalias{taylor2012_VC1}.

    One potential complication in this comparison arises from the complex spatial structure of the VC1 and WAVES South regions, both of which contain varying populations at distances ranging from 17 to 32~Mpc. However, the disparity remains evident even when restricting the analysis to the main cluster body at 17~Mpc. For this subset the detection fraction for VCC galaxies ($\frac{N_{\HI{},VCC, 17~\rm{Mpc}}}{N_{VCC, 17~\rm{Mpc}}}$) is 11\% in WAVES South compared to 17\% in VC1. This result is significant given that WAVES South is heavily dominated by the 17~Mpc objects, accounting for approximately 71\% of its VCC population. These objects, on the other hand, make up less than half of the VC1 sample. Therefore, despite having a higher concentration of galaxies at the distance where \HI{} is more likely to be detected, WAVES South still exhibits a notable deficit of gas-bearing systems.

    Finally, the census of "dark" \HI{} clouds arguably also provides support for interpreting WAVES South as more relaxed. While VC1 hosts eight isolated dark clouds, WAVES South contains only one truly dark candidate, WCS~54. Crucially, WCS~54 is not isolated but is connected to the galaxy WCS~40 (VCC~952) via an \HI{} bridge. The existence of WCS~54, combined with the low \HI{} deficiency of its likely parent WCS~40, shown in Fig.\ref{HI_def_dist_fig}, suggests we have captured a recent arrival to the cluster caught in the early stages of stripping within an otherwise evolved region. If indeed WAVES South is more evolved than VC1, then the lack of isolated clouds might be expected, with such features being disrupted over time as they move through the cluster (but see \citealt{bellazzini2018alone,calura2020hydrodynamic} for simulations showing that BBs may be long lived). Thus we would expect to find more clouds in regions of greater ongoing galaxy infall such as VC1. Capturing a single cloud in formation in WAVES South does not contradict this, but fully evaluating this hypothesis requires deep data of other parts of the cluster, such as WAVES North.

    The picture from \HI{} deficiency, on the other hand, is less clear. WAVES South is if anything slightly less deficient than VC1, which is not what we might expect in a more evolved area. However the statistical significance of this is low due to the low sample size and, as in \citet{taylor2020faint}, the \HI{} deficiency is not an indication of ongoing stripping.

    \subsection{Dark clouds and blue blobs}
    The only dark cloud candidate in WAVES South WCS~54 is notably distinct from the isolated dark clouds found in VC1. The presence of an \HI{} bridge connecting it to WCS~40, alongside compressed contours and a secondary tail, strongly suggests a very recent ram-pressure stripping formation mechanism. While WAVES South lacks the larger isolated dark cloud population seen in VC1, it hosts several other peculiar systems, many of which are associated with the recently cataloged BBs \citepalias{dey2025citizen}. The blue blob population is remarkably prominent in this part of the cluster. Combined, the WAVES South and VC1 regions contain almost half of the \citetalias{dey2025citizen} catalog, with both regions hosting exactly five BBs detected in \HI{} and four that are undetected. Considering that WAVES South is likely more relaxed and has fewer dark clouds than VC1, the fact that it contains the same number of BBs is surprising. This suggests that the number of BBs is not directly tied to the number of ongoing or recent stripping events.

    The VC1 region exhibits a nearly identical distribution of detected BBs. While the VC1 population similarly lacks a clear spatial trend, there is a notable association with objects previously classified as dark clouds. Specifically, BC17 and BC25 have been identified as the optical counterparts to the clouds AGESVC1~266 and AGESVC1~274, respectively (\citetalias{dey2025citizen}; \citealt{minchin2026high}). This fact, taken together with the isolated nature of both populations, hints at a potential evolutionary connection between "dark" clouds and BBs, where the clouds represent a precursor stage to the star-forming blue blob phase. If BBs represent a long-lived stage in the evolution of dark clouds, this could explain why more such clouds are found in the less-relaxed VC1 area compared to WAVES South. In this scenario, stripped \HI{} might exist in a dark phase until local densities or environmental pressures trigger a late-stage bout of star formation, effectively transforming the dark cloud into a blue blob. We caution however, that this assumption is reliant on very small number statistics and a proper evaluation requires AGES-depth sensitivity of a much larger area of the cluster. Such a prospect will ultimately be investigated with FAST and/or possibly with LOFAR and MeerKAT via the ViCTORIA project \citep{de2025victoria}.

    For the moment we must depend on the analysis of individual objects. In WAVES South, three of the \HI{}-detected BBs are associated with peculiar or highly disturbed systems rather than regular galaxies. These objects typically exhibit complex \HI{} profiles without clear directional patterns or tails, which would fit the picture of ram-pressure dwarfs, as predicted by \citet{jones2022young}. In contrast, WCS~14 (BC33) and WCS~46 (BC40) resemble more typical dwarf galaxies, characterized by compact \HI{} profiles and moderate deficiencies. These properties indicate they are indeed stable, possibly self-bound, following a recent ram-pressure stripping episode, although further research is needed. Alternatively, \citetalias{dey2025citizen} classifies these two systems as "rank 2" BBs, which might be contaminated by regular galaxies that share similar properties with BBs, but are not necessarily products of stripped gas. As in the case of the dark clouds analyzed in \citet{minchin2026high}, the analysis of these objects is further complicated as even these small populations are likely not homogeneous, with a range of formation mechanisms at work.

    Furthermore, based on the observed \HI{} deficiency, approximately $3\times10^{10}~\rm{M_{\odot}}$ is "missing" from the cluster galaxies in WAVES South. Even if all the extragalactic gas detected in this region$-$including the combined mass of the BBs (with both \HI{} and stellar components), WCS~54, the star-forming cloud WCS~11, and the massive AV7 complex (which accounts for more than half of the entire mass)$-$had originated from these galaxies, it would account for less than 7\% of this total deficit. The BBs with their \HI{} components alone contribute less than 3\%, a value compatible with a ram-pressure origin where the vast majority of stripped gas eventually mixes with or, more likely, evaporates into the hot ICM. 

    While we did not find any parent association, this possibility cannot definitively be ruled out; if BBs are indeed relatively long-lived systems, their formation could have occurred in the distant past. This interpretation is consistent with numerical simulations suggesting that such fragments can persist over significant timescales \citep{bellazzini2018alone,calura2020hydrodynamic}. Nonetheless, our data suggests that the BBs in WAVES South are likely not the result of recent gas removal events.

    \subsection{Gas-loss fueled evolution}\label{gas_loss_discussion}
    In the VC1 region, \citetalias{taylor2012_VC1} identified several ETGs, including three dwarf ellipticals with unexpectedly high gas fractions. At first glance, these findings appeared to conflict with the evolutionary sequence proposed by \citet{boselli2008origin}, wherein dIrr galaxies are rapidly stripped of their gas via ram-pressure stripping, subsequently quenching and transforming into gas-poor dEs. \citetalias{taylor2012_VC1} suggested that these gas-rich dEs might represent recent cluster arrivals or instances of morphological misclassification. Apart from VCC~1964 (WCS~11), which we present separately in \citet{taylor2026ultra}, we were able to detect two ETGs in \HI{} within WAVES South: the dE WCS~47 (VCC~21) and the dwarf spheroidal WCS~51 (VCC~651).

    As shown in the \HI{}-to-stellar mass ratio diagram (Fig.~\ref{gasfrac_fig}), both galaxies exhibit \HI{}-to-stellar mass fractions significantly below the median for our \HI{}-detected sample. This observed gas-poor state is consistent with the late stages of a stripping-induced transformation. If these objects are indeed transitioning dIrrs that have been misclassified due to their increasingly smooth optical profiles, their implied \HI{} deficiencies would exceed $DEF_{\HI{}}>1$.

    The presence of a residual, low-mass \HI{} component in an otherwise smooth, early-type morphology suggests that we are observing these systems during the final phase of their transition into quenched ETGs. While our current sample size is small, the characteristics of WCS~47 and WCS~51 align with the gas-stripping and morphological transformation model of \citet{boselli2008origin}, reflecting a more advanced evolutionary state compared to the gas-rich ETGs reported in VC1. 

    Conversely, we identified two groups of true LTGs lacking \HI{}. As discussed in Sect.~\ref{sec_morphology}, the similarity of the SDSS/WISEA objects is consistent with the scenario proposed in \citet{kleiner2023meerkat}, namely, that these are the result of very recent RPS, with the galaxies not having yet evolved significantly in color and structure since their gas-loss phase (see also \citealt{taylor2026ultra}). We note here that the \HI{} non-detected VCC galaxies could follow a similar evolutionary path, however, as they are typically an order of magnitude more massive, the effects of ram pressure would be expected to be more gradual.

    \subsection{Stacking}
    To maximize our sensitivity to low-level emission, we performed a stacking analysis using all VCC galaxies listed in the GOLDMine catalog that lacked individual \HI{} detections. We constructed various subsamples based on the survey footprint (WAVES South, VC1, and VC2), morphology, and distance. While no new \HI{} signals were identified in the resulting stacks, these results allowed us to establish rigorous upper limits on the neutral gas content for several galaxy populations. 

    The most significant constraint was derived from a stack of 157 galaxies across three regions. This stack reached a noise level of rms$=0.080~\rm{mJy}$, which corresponds to a $3\sigma$ \HI{} upper limit of $1.26\times10^6~\rm{M_{\odot}}$. This is further improvement, albeit slight, over \citetalias{taylor2012_VC1}, where they achieved rms of $0.088~\rm{mJy}$ in their deepest stack. The lack of detection is consistent with the findings of previous studies and, alongside the findings of \citet{conselice2003galaxy}, reinforces the conclusion that ETGs in the Virgo cluster are essentially devoid of neutral gas. 

    This persistent lack of detection suggests that once a galaxy's gas density reaches a critically low threshold (likely comparable to the density of the surrounding ICM) ram-pressure stripping becomes highly efficient at clearing the remaining reservoir. Alternatively, any remaining low-density gas may become susceptible to rapid ionization. The persistence of these non-detections across all regions, including the infalling populations of VC1 where the ICM is presumably less dense, indicates that no "hidden" gas reservoir exists below our current individual detection limits.

    Similar trends were observed by \citet{healy2021h} in the Coma cluster. They reported a significant drop in the \HI{} mass function (HIMF) below $\sim10^8~\rm{M_{\odot}}$. Consistently with our results, their study yielded no new detection below their nominal threshold when stacking galaxies within the cluster core (although they did successfully recover a stacked signal in the cluster outskirts). In Virgo, we see detections continuous in mass down to $\sim10^7~\rm{M_{\odot}}$ (see Fig.~\ref{mass_width_fig}), below which there is a sharp decrease in \HI{} detections that remains unrecoverable even through our deepest stacks (see also \citealt{davies2011arecibo} for discussion on Virgo HIMF). Recent ultra-deep, direct observations of the Fornax cluster by \citet{kleiner2025meerkat} have also revealed an abrupt collapse of the HIMF below $\sim10^7~\rm{M_{\odot}}$, a threshold that is remarkably similar to the one we inferred for Virgo. The higher mass cut-off observed in the Coma HIMF is possibly indicative of its denser and hotter ICM, creating a harsher stripping environment than Virgo. Interestingly, although Fornax contains an ICM roughly half as dense as that of Virgo, it exhibits a similar HIMF cut-off, possibly driven by its higher galaxy density and consequently greater role of tidal interactions. 

    While the details of the mechanisms causing the gas loss are still not fully understood, the environment does appear to be the primary driver of this gas depletion. Unlike field ETGs, where stacking successfully recovers \HI{} signals and dwarf systems show no sharp ionization threshold \citep{fabello2011alfalfa,ianjamasimanana2018smooth,deshev2022arecibo}, the stark contrast in Virgo suggests the cluster environment imposes a critical density limit on \HI{} survival. In such an extreme environment, the survival of apparently long-lived, extremely dim and dark gas clouds becomes all the more intriguing. Understanding how such low-mass features can survive when more massive galaxies end up efficiently stripped will require additional observational data and numerical simulations. 

    \section*{Data availability} 
    Tables \ref{HI_properties_table}, \ref{WCS_optical_table} and \ref{nonHI_properties_table} are available in electronic form at the CDS via anonymous ftp to cdsarc.u-strasbg.fr (130.79.128.5) or via http://cdsweb.u-strasbg.fr/cgi-bin/qcat?J/A+A/. The WAVES South data cube is available in its entirety on request to the corresponding author. All other data sources used in this work are public.

\begin{acknowledgements}
    We thank the anonymous referee for thorough and constructive comments that significantly improved the quality of this work.

    This work was supported by the institutional project RVO: 67985815, the Czech Ministry of Education, Youth and Sports from the large infrastructures for Research, Experimental Development and Innovations project LM 2015067, and the Charles University project GA UK No. 376425. 

    This research has made use of the NASA/IPAC Extragalactic Database (NED), which is operated by the Jet Propulsion Laboratory, California Institute of Technology, under contract with the National Aeronautics and Space Administration.

    This research has made use of the GOLDMine Database and the DESI Legacy Survey data.
    
    This work has made use of the SDSS. Funding for the Sloan Digital Sky Survey V has been provided by the Alfred P. Sloan Foundation, the Heising-Simons Foundation, the National Science Foundation, and the Participating Institutions. SDSS acknowledges support and resources from the Center for High-Performance Computing at the University of Utah. SDSS telescopes are located at Apache Point Observatory, funded by the Astrophysical Research Consortium and operated by New Mexico State University, and at Las Campanas Observatory, operated by the Carnegie Institution for Science. The SDSS web site is \url{www.sdss.org}. SDSS is managed by the Astrophysical Research Consortium for the Participating Institutions of the SDSS Collaboration, including the Carnegie Institution for Science, Chilean National Time Allocation Committee (CNTAC) ratified researchers, Caltech, the Gotham Participation Group, Harvard University, Heidelberg University, The Flatiron Institute, The Johns Hopkins University, L'Ecole polytechnique f\'{e}d\'{e}rale de Lausanne (EPFL), Leibniz-Institut f\"{u}r Astrophysik Potsdam (AIP), Max-Planck-Institut f\"{u}r Astronomie (MPIA Heidelberg), Max-Planck-Institut f\"{u}r Extraterrestrische Physik (MPE), Nanjing University, National Astronomical Observatories of China (NAOC), New Mexico State University, The Ohio State University, Pennsylvania State University, Smithsonian Astrophysical Observatory, Space Telescope Science Institute (STScI), the Stellar Astrophysics Participation Group, Universidad Nacional Aut\'{o}noma de M\'{e}xico, University of Arizona, University of Colorado Boulder, University of Illinois at Urbana-Champaign, University of Toronto, University of Utah, University of Virginia, Yale University, and Yunnan University.  

\end{acknowledgements}

\bibpunct{(}{)}{;}{a}{}{,}

\bibliographystyle{aa}
\bibliography{bibliogaphy}

\begin{appendix}
\onecolumn

\begingroup
\begin{table*}[ht]
    \section{Catalog of \HI{} sources in WAVES South}
    \caption{Catalog of \HI{} properties of the WCS objects. Coordinates of \HI{} are in J2000 with errors in the brackets as calculated by \textsc{Miriad}. Coordinates without brackets indicate that the position fitting failed. In those cases, they were reverted to the initial by-eye estimates. Flux is the total integrated flux of the line. S/N column shows the integrated signal-to-noise value calculated by Eq.~\ref{SN_equation}.}             
    \label{HI_properties_table}      
    \centering     
    \small
    \setlength{\tabcolsep}{5pt}
    \centerline{
    \begin{tabular}{l c c c c c c c c c c}     
    \hline\hline       
 
    WCS     & R.A.             & Dec.              & Vel.      &  W50      & W20       & Flux           & $rms$ & Dist.& $M_{\HI{}}$& $\rm{S/N}_{int}$ \\
    Name    &                  &                   & [$\rm{km\,s^{-1}}$]  & [$\rm{km\,s^{-1}}$] & [$\rm{km\,s^{-1}}$] & [$\rm{Jy\,km\,s^{-1}}$] & [$\rm{mJy\,beam^{-1}}$] & [$\rm{Mpc}$] & [$\rm{logM_\odot}$] & \\
    \hline

    WCS 1   & 12:33:37.9(0.7)  & +09:10:38.94(10)  &  2332(2)  &  205(5)   &  251(7)   &  6.517(0.334)  &  0.7  &  17.0  &  8.65      &  145.4     \\
    WCS 2   & 12:36:38.0(0.9)  & +11:11:58.63(15)  &  2240(3)  &  288(7)   &  331(10)  &  6.278(0.439)  &  1.9  &  17.0  &  8.63      &  43.5      \\
    WCS 3   & 12:36:10.5(0.7)  & +09:54:52.86(10)  &  2055(2)  &  73(5)    &  113(7)   &  2.222(0.189)  &  0.8  &  17.0  &  8.18      &  72.7      \\
    WCS 4   & 12:35:13.8(0.7)  & +10:26:08.53(10)  &  1082(2)  &  82(4)    &  105(7)   &  1.153(0.122)  &  0.8  &  17.0  &  7.90      &  35.6      \\
    WCS 5   & 12:32:32.7(0.7)  & +09:10:27.99(10)  &  1158(2)  &  69(4)    &  100(6)   &  1.793(0.166)  &  0.8  &  17.0  &  8.09      &  60.3      \\
    WCS 6   & 12:46:08.4(0.8)  & +10:09:46.40(11)  &  1506(2)  &  50(4)    &  66(6)    &  0.767(0.111)  &  1.0  &  17.0  &  7.72      &  24.3      \\
    WCS 7   & 12:48:00.2(0.9)  & +10:58:17.26(10)  &  1152(2)  &  56(4)    &  91(6)    &  3.715(0.327)  &  0.9  &  17.0  &  8.40      &  123.3     \\
    WCS 8   & 12:46:14.9(0.8)  & +10:11:32.70(13)  &  1147(4)  &  37(7)    &  56(11)   &  0.493(0.130)  &  1.6  &  17.0  &  7.53      &  11.3      \\
    WCS 9   & 12:46:41.0(0.9)  & +10:21:49.10(14)  &  1590(3)  &  29(7)    &  70(10)   &  0.359(0.064)  &  0.7  &  17.0  &  7.39      &  21.3      \\
    WCS 10  & 12:48:40.2(0.8)  & +10:44:53.39(11)  &  1614(3)  &  41(6)    &  53(9)    &  0.264(0.068)  &  0.8  &  17.0  &  7.26      &  11.5      \\
    WCS 11  & 12:43:18.0(1.4)  & +08:55:09.10(14)  &  1429(3)  &  25(6)    &  40(9)    &  0.465(0.133)  &  1.9  &  17.0  &  7.50      &  10.9      \\
    WCS 12  & 12:35:01.6(0.8)  & +09:10:26.31(15)  &  1291(3)  &  50(6)    &  62(9)    &  0.290(0.072)  &  0.8  &  17.0  &  7.30      &  11.5      \\
    WCS 13  & 12:38:07.7(0.8)  & +10:09:25.82(24)  &  1150(2)  &  25(5)    &  39(7)    &  0.230(0.057)  &  0.8  &  17.0  &  7.20      &  12.9      \\
    WCS 14  & 12:30:34.1(0.9)  & +10:15:49.96(11)  &  1108(2)  &  23(4)    &  39(6)    &  0.349(0.067)  &  0.8  &  17.0  &  7.38      &  20.3      \\
    WCS 15  & 12:21:28.3(0.8)  & +10:29:56.21(13)  &  997(4)   &  27(9)    &  42(13)   &  0.119(0.052)  &  0.8  &  17.0  &  6.91      &  6.4       \\
    WCS 16  & 12:28:53.3(0.9)  & +09:25:02.11(11)  &  1055(3)  &  104(6)   &  116(8)   &  0.697(0.115)  &  0.9  &  23.0  &  7.94      &  17.0      \\
    WCS 17  & 12:34:30.3(0.7)  & +09:09:12.10(10)  &  448(2)   &  89(4)    &  111(7)   &  1.360(0.132)  &  0.8  &  17.0  &  7.97      &  40.3      \\
    WCS 18  & 12:29:41.1(0.7)  & +09:42:45.05(10)  &  538(2)   &  123(4)   &  143(6)   &  1.325(0.113)  &  0.6  &  23.0  &  8.22      &  44.5      \\
    WCS 19  & 12:31:25.6(0.7)  & +09:18:49.68(10)  &  485(3)   &  60(6)    &  104(9)   &  0.945(0.103)  &  0.7  &  23.0  &  8.07      &  39.0      \\
    WCS 20  & 12:30:30.1       & +09:31:23.83      &  499(8)   &  98(15)   &  322(23)  &  1.510(0.148)  &  0.9  &  23.0  &  8.28      &  37.9      \\
    WCS 21  & 12:31:18.9(0.7)  & +09:28:36.61(11)  &  606(3)   &  52(6)    &  86(9)    &  0.826(0.116)  &  1.0  &  23.0  &  8.01      &  25.6      \\
    WCS 22  & 12:29:23.3(1.0)  & +09:57:56.14(19)  &  792(10)  &  93(20)   &  164(30)  &  0.487(0.120)  &  1.0  &  23.0  &  7.78      &  11.3      \\
    WCS 23  & 12:40:02.2(0.8)  & +09:25:38.85(12)  &  1394(4)  &  50(8)    &  69(12)   &  0.188(0.056)  &  0.7  &  17.0  &  7.11      &  8.5       \\
    WCS 24  & 12:45:09.0(0.8)  & +10:19:49.60(24)  &  2203(8)  &  48(15)   &  86(23)   &  0.177(0.065)  &  0.8  &  17.0  &  7.08      &  7.2       \\
    WCS 25  & 12:15:45.2(0.7)  & +10:42:01.18(10)  &  1985(1)  &  209(3)   &  226(4)   &  12.678(0.616) &  1.1  &  17.0  &  8.94      &  178.3     \\
    WCS 26  & 12:27:21.4(0.7)  & +10:52:06.19(10)  &  929(1)   &  159(3)   &  173(4)   &  7.420(0.393)  &  0.7  &  17.0  &  8.70      &  188.0     \\
    WCS 27  & 12:22:41.9(0.7)  & +09:20:07.34(10)  &  1257(1)  &  309(3)   &  324(4)   &  8.178(0.400)  &  0.9  &  23.0  &  9.01      &  115.6     \\
    WCS 28  & 12:25:50.1(0.7)  & +10:27:42.02(10)  &  1105(2)  &  127(3)   &  148(5)   &  5.961(0.369)  &  0.8  &  23.0  &  8.87      &  147.8     \\
    WCS 29  & 12:15:30.3(0.7)  & +09:34:56.39(10)  &  600(2)   &  200(4)   &  225(6)   &  5.602(0.307)  &  0.9  &  17.0  &  8.58      &  98.4      \\
    WCS 30  & 12:14:35.5(0.7)  & +09:11:54.71(10)  &  1794(1)  &  123(3)   &  139(4)   &  3.963(0.256)  &  0.8  &  32.0  &  8.98      &  99.9      \\
    WCS 31  & 12:10:56.9(0.7)  & +09:13:14.55(10)  &  2246(2)  &  130(4)   &  145(7)   &  1.466(0.149)  &  0.9  &  32.0  &  8.55      &  31.9      \\
    WCS 32  & 12:12:00.7(0.7)  & +10:23:59.49(10)  &  1636(2)  &  77(4)    &  91(5)    &  0.988(0.105)  &  0.7  &  17.0  &  7.83      &  36.0      \\
    WCS 33  & 12:17:03.9(0.7)  & +10:00:15.74(10)  &  1188(2)  &  57(3)    &  77(5)    &  3.866(0.340)  &  0.9  &  17.0  &  8.42      &  127.2     \\
    WCS 34  & 12:15:54.8(0.7)  & +09:39:07.21(10)  &  2226(1)  &  26(3)    &  40(4)    &  4.036(0.516)  &  1.0  &  17.0  &  8.44      &  177.0     \\
    WCS 35  & 12:12:45.9(0.7)  & +10:52:15.04(10)  &  384(2)   &  248(4)   &  282(6)   &  24.760(1.079) &  0.7  &  17.0  &  9.23      &  502.2     \\
    WCS 36  & 12:26:41.1       & +08:56:59.90      &  1278(2)  &  84(3)    &  102(5)   &  4.133(0.323)  &  1.2  &  23.0  &  8.71      &  84.0      \\
    WCS 37  & 12:28:40.7(0.7)  & +09:15:40.38(10)  &  866(2)   &  209(5)   &  228(7)   &  2.322(0.185)  &  0.9  &  23.0  &  8.46      &  39.9      \\
    WCS 38  & 12:27:15.5       & +09:22:51.85      &  439(2)   &  58(5)    &  97(7)    &  2.128(0.197)  &  0.9  &  23.0  &  8.42      &  69.4      \\
    WCS 39  & 12:25:22.1(0.7)  & +10:00:58.12(10)  &  970(2)   &  271(5)   &  291(7)   &  2.380(0.176)  &  0.8  &  23.0  &  8.47      &  40.4      \\
    WCS 40  & 12:26:56.2(0.7)  & +09:53:19.12(10)  &  990(2)   &  51(5)    &  83(7)    &  1.133(0.124)  &  0.8  &  23.0  &  8.15      &  44.3      \\
    WCS 41  & 12:15:04.2(0.7)  & +09:45:25.18(10)  &  2199(2)  &  109(4)   &  124(6)   &  1.238(0.128)  &  0.8  &  17.0  &  7.93      &  33.1      \\
    WCS 42  & 12:19:00.5(0.9)  & +08:55:12.52(13)  &  2468(3)  &  162(6)   &  181(8)   &  2.098(0.238)  &  1.5  &  32.0  &  8.71      &  24.6      \\
    WCS 43  & 12:16:32.7(0.7)  & +10:12:35.29(10)  &  2072(2)  &  55(5)    &  75(7)    &  0.847(0.117)  &  1.0  &  17.0  &  7.76      &  25.5      \\
    WCS 44  & 12:14:12.6(0.7)  & +08:52:35.17(21)  &  1934(2)  &  84(4)    &  99(7)    &  3.291(0.431)  &  3.3  &  32.0  &  8.90      &  24.3      \\
    WCS 45  & 12:15:28.0(0.7)  & +10:30:33.66(10)  &  2000(3)  &  53(6)    &  91(9)    &  0.746(0.091)  &  0.7  &  17.0  &  7.71      &  32.7      \\
    WCS 46  & 12:21:11.7(0.7)  & +10:37:45.63(10)  &  2613(2)  &  39(5)    &  57(7)    &  0.464(0.078)  &  0.8  &  17.0  &  7.50      &  20.8      \\
    WCS 47  & 12:10:23.2(0.7)  & +10:11:35.90(11)  &  502(3)   &  82(6)    &  91(10)   &  0.591(0.148)  &  1.4  &  17.0  &  7.61      &  10.4      \\
    WCS 48  & 12:22:04.7(0.7)  & +09:02:43.90(10)  &  1053(4)  &  310(7)   &  324(11)  &  1.545(0.194)  &  1.0  &  23.0  &  8.29      &  19.6      \\
    WCS 49  & 12:25:35.0(1.0)  & +10:35:44.89(14)  &  1719(5)  &  24(10)   &  56(16)   &  0.075(0.035)  &  0.7  &  23.0  &  6.97      &  4.9       \\
    WCS 50  & 12:28:15.9(0.7)  & +09:26:17.87(12)  &  431(5)   &  101(9)   &  124(14)  &  0.517(0.107)  &  0.9  &  23.0  &  7.81      &  12.8      \\
    WCS 51  & 12:23:33.6(1.0)  & +09:31:22.70(15)  &  964(4)   &  24(8)    &  34(13)   &  0.137(0.072)  &  1.2  &  23.0  &  7.23      &  5.2       \\
    WCS 52  & 12:27:35.7(1.0)  & +10:00:20.76(11)  &  881(3)   &  51(7)    &  61(10)   &  0.215(0.078)  &  1.1  &  23.0  &  7.43      &  6.1       \\
    WCS 53  & 12:27:29.2(0.9)  & +09:20:10.40(13)  &  1719(5)  &  33(10)   &  55(16)   &  0.208(0.080)  &  1.1  &  23.0  &  7.41      &  7.4       \\
    WCS 54  & 12:27:35.7(0.9)  & +09:51:33.31(16)  &  954(6)   &  68(11)   &  85(17)   &  0.292(0.099)  &  1.0  &  23.0  &  7.56      &  7.9       \\
    WCS 55  & 12:25:32.5       & +11:10:45.88      &  903(4)   &  85(9)    &  92(13)   &  0.390(0.200)  &  2.4  &  17.0  &  7.42      &  3.9       \\
    WCS 56  & 12:14:55.05(0.8) & +09:40:31.63(11)  &  1722(9)  &  34(17)   &  73(26)   &  0.142(0.066)  &  0.9  &  17.0  &  6.99      &  6.0       \\
    \hline                  
    \end{tabular}
    }
\end{table*}
\endgroup
\pagebreak

\begin{table*}[ht]
    \section{Optical properties of the \HI{}-detected galaxies}
    \caption{Optical properties of the WCS galaxies. Type shows the morphological type using the GOLDMine system and value, unless stated otherwise. CL shows confidence levels, as defined in Sect.~\ref{section_data_extraction}. The $m_g$ and $m_i$ magnitudes are given by the SDSS. $M_{\rm{\HI{} }}/M_{\star}$ shows the \HI{}-to-stellar-mass ratio. $a$ shows the GOLDMine optical diameter, unless stated otherwise. $DEF_{\rm{\HI{}}}$ shows the \HI{} deficiency. This table does not include objects with no galactic counterpart or pairs of galaxies with indistinguishable \HI{} emission.}             
    \label{WCS_optical_table}      
    \small
    \centering
    \centerline{
    \begin{tabular}{l l c c c c c c c c c c }     
    \hline\hline       
    WCS      &  Other             &   R.A.		    &   Dec.	      &  Type   &   CL	  &  $m_g$   &   $m_i$  &  $g-i$  & $M_{\rm{\HI{}}}/M_{\star}$ &      $a$    &   $DEF_{\rm{\HI{}}}$  \\
    Name     &  Name              &                 &                 &         &         &          &          &         &                       &    [arcsec] &                \\
    \hline
    WCS 1	 &  VCC 1516	      &   12:33:39.71	&   +09:10:30.1	  &  6  	&   0	  &  12.75	 &   11.76	&  0.98	  &  0.11	              &      242.4	&   0.76         \\
    WCS 3	 &  VCC 1654	      &   12:36:10.58	&   +09:55:21.2	  &  12 	&   0	  &  15.90	 &   15.52	&  0.38	  &  3.22	              &      51.0	&   0.24         \\
    WCS 4\tablefootmark{b}	 &  VCC 1605	      &   12:35:13.91	&   +10:25:53.3	  &  9  	&   0	  &  17.05	 &   16.43	&  0.63	  &  2.59	              &      60.0	&   0.68         \\
    WCS 5	 &  VCC 1437	      &   12:32:33.50	&   +09:10:25.2	  &  17  	&   0	  &  14.71	 &   14.07	&  0.64	  &  0.45	              &      35.4	&   0.10          \\
    WCS 6	 &  VCC 2034	      &   12:46:08.28	&   +10:09:51.5	  &  12  	&   0	  &  17.47	 &   16.75	&  0.72	  &  2.00	              &      46.8	&   0.67         \\
    WCS 8	 &  VCC 2037	      &   12:46:15.31	&   +10:12:26.6	  &  16  	&   0	  &  16.37	 &   15.85	&  0.52	  &  0.78	              &      52.8	&   0.92         \\
    WCS 10\tablefootmark{a}		 &  NA	              &   12:48:40.34	&   +10:44:15.8	  &  12     &   1	  &  18.50	 &   18.70	&  -0.20	  &  18.46	              &      18.3   &   0.51         \\
    WCS 11\tablefootmark{b}	 &  VCC 1964	      &   12:43:18.12	&   +08:55:18.8	  &  -1  	&   1	  &  17.39	 &   16.94	&  0.45	  &  2.24	              &      31.2	&   0.80          \\
    WCS 12	 &  VCC 1596	      &   12:35:00.96	&   +09:11:09.9	  &  12  	&   0	  &  17.56	 &   17.17	&  0.39	  &  1.90	              &      21.0	&   0.56         \\
    WCS 13\tablefootmark{b}	 &  VCC 1744	      &   12:38:06.89	&   +10:09:56.0	  &  17  	&   0	  &  16.41	 &   16.25	&  0.15	  &  0.95	              &      30.6	&   0.90          \\
    WCS 14\tablefootmark{b}	 &  VCC 1295	      &   12:30:32.30	&   +10:15:40.5	  &  20  	&   0	  &  18.06	 &   17.23	&  0.83	  &  1.26	              &      22.8	&   0.58         \\
    WCS 15	 &  VCC 476	          &   12:21:28.92	&   +10:29:08.7	  &  12  	&   1	  &  18.07	 &   17.41	&  0.65	  &  0.64	              &      21.6	&   0.97         \\
    WCS 16	 &  VCC 1141	      &   12:28:54.95	&   +09:25:16.0	  &  17  	&   0	  &  16.22	 &   15.56	&  0.66	  &  0.67	              &      27.6	&   0.28         \\
    WCS 17	 &  VCC 1566	      &   12:34:30.99	&   +09:09:20.1	  &  9  	&   0	  &  15.19	 &   14.6	&  0.58	  &  0.61	              &      69.6	&   0.65         \\
    WCS 22	 &  VCC 1179	      &   12:29:22.66	&   +09:59:17.8	  &  16  	&   0	  &  15.46	 &   14.98	&  0.49	  &  0.36	              &      69.6	&   1.03         \\
    WCS 23\tablefootmark{a}	 &  NA	              &   12:40:03.09	&   +09:24:49.0	  &  12     &   1	  &  18.15	 &   17.84	&  0.31	  &  2.60	              &      11.5	&   0.37         \\
    WCS 24	 &  VCC 2015	      &   12:45:11.83	&   +10:19:27.5	  &  17  	&   0	  &  16.08	 &   15.62	&  0.46	  &  0.25	              &      30.6	&   1.01         \\
    WCS 25	 &  VCC 162	          &   12:15:46.27	&   +10:41:57.4	  &  9  	&   0	  &  14.94	 &   14.27	&  0.67	  &  3.68	              &      175.2	&   0.27         \\
    WCS 26	 &  VCC 995	          &   12:27:22.20	&   +10:52:00.1	  &  7  	&   0	  &  15.43	 &   14.92	&  0.51	  &  5.06	              &      91.8	&   0.40          \\
    WCS 27	 &  VCC 576	          &   12:22:42.24	&   +09:19:56.9	  &  6  	&   0	  &  13.49	 &   12.19	&  1.31	  &  0.12	              &      148.8	&   0.28         \\
    WCS 28	 &  VCC 849	          &   12:25:50.67	&   +10:27:32.6	  &  6  	&   0	  &  13.22	 &   12.55	&  0.67	  &  0.35	              &      130.8	&   0.34         \\
    WCS 29	 &  VCC 152	          &   12:15:30.50	&   +09:35:05.6	  &  8  	&   0	  &  13.23	 &   12.03	&  1.20	  &  0.09	              &      117.6	&   0.37         \\
    WCS 30	 &  VCC 117	          &   12:14:35.67	&   +09:11:59.2	  &  12  	&   0	  &  16.50	 &   16.06	&  0.44	  &  8.57	              &      38.4	&   -0.34        \\
    WCS 31	 &  VCC 31	          &   12:10:57.10	&   +09:13:09.9	  &  20  	&   0	  &  15.64	 &   14.99	&  0.65	  &  0.86	              &      37.8	&   0.08         \\
    WCS 32\tablefootmark{a}	 &  VCCA 046	      &   12:12:01.21	&   +10:23:54.4	  &  5   	&   0	  &  17.73	 &   17.35	&  0.38	  &  7.78	              &      33.0	&   0.32         \\
    WCS 33	 &  VCC 217	          &   12:17:04.23	&   +10:00:20.1	  &  12  	&   0	  &  17.37	 &   16.86	&  0.51	  &  15.81	              &      102.6	&   0.29         \\
    WCS 34\tablefootmark{b}	 &  VCC 169           &   12:15:56.32	&   +09:38:57.6	  &  12  	&   0	  &  16.39	 &   16.07	&  0.33	  &  10.62	              &      51.0	&   0.18         \\
    WCS 35	 &  VCC 66	          &   12:12:46.33	&   +10:51:54.9	  &  7  	&   0	  &  12.71	 &   11.76	&  0.95	  &  0.45	              &      321.0	&   0.36         \\
    WCS 37	 &  VCC 1118	      &   12:28:40.52	&   +09:15:34.1	  &  7  	&   0	  &  13.14	 &   12.24	&  0.90	  &  0.07	              &      117.6	&   0.68         \\
    WCS 38	 &  VCC 979	          &   12:27:11.84	&   +09:25:12.3	  &  3  	&   0	  &  12.38	 &   11.52	&  0.85	  &  0.04	              &      259.8	&   1.22         \\
    WCS 39	 &  VCC 792	          &   12:25:22.16	&   +10:01:00.5	  &  4  	&   0	  &  12.92	 &   11.72	&  1.20	  &  0.03	              &      211.2	&   1.04         \\
    WCS 40	 &  VCC 952	          &   12:26:55.72	&   +09:52:56.0	  &  12  	&   0	  &  16.86	 &   16.65	&  0.21	  &  6.11	              &      46.8	&   0.41         \\
    WCS 41	 &  VCC 130	          &   12:15:04.01	&   +09:45:13.2	  &  17  	&   0	  &  16.77	 &   16.33	&  0.44	  &  3.43	              &      37.8	&   0.30          \\
    WCS 42	 &  VCC 318	          &   12:19:03.84	&   +08:51:29.1	  &  8  	&   0	  &  14.88	 &   14.70	&  0.17	  &  2.01	              &      102.6	&   0.56         \\
    WCS 43\tablefootmark{a}	 &  VCCA 059	      &   12:16:34.03	&   +10:12:22.8	  &  12     &   0	  &  17.39	 &   17.15	&  0.24	  &  6.99	              &      11.5	&   -0.29        \\
    WCS 44\tablefootmark{a,b}  	 &  VCCA 052	      &   12:14:13.79	&   +08:54:29.8	  &  5   	&   0	  &  17.22	 &   17.05	&  0.18	  &  27.10	              &      37.8	&   -0.27        \\
    WCS 45\tablefootmark{a}	 &  VCCA 056	      &   12:15:27.34	&   +10:30:44.4	  &  12     	&   0	  &  17.48	 &   17.24	&  0.25	  &  6.53	              &      14.1	&   -0.10         \\
    WCS 46\tablefootmark{a}	 &  VCCA 079	      &   12:21:13.23	&   +10:37:32.0	  &  12  	&   0	  &  18.76	 &   18.41	&  0.35	  &  10.04	              &      22.3	&   0.40          \\
    WCS 47	 &  VCC 21	          &   12:10:23.09	&   +10:11:18.8	  &  -3  	&   0	  &  14.76	 &   14.12	&  0.64	  &  0.16	              &      72.0	&   1.03         \\
    WCS 48	 &  VCC 524	          &   12:22:05.67	&   +09:02:37.0	  &  6  	&   0	  &  12.97	 &   11.63	&  1.34	  &  0.01	              &      237.0	&   1.30          \\
    WCS 49	 &  VCC 825	          &   12:25:39.34	&   +10:35:02.0	  &  12  	&   0	  &  16.27	 &   15.62	&  0.64	  &  0.08	              &      60.0	&   1.74         \\
    WCS 50	 &  VCC 1086	      &   12:28:15.93	&   +09:26:10.7	  &  18  	&   0	  &  13.64	 &   12.50	&  1.14	  &  0.01	              &      192.0	&   1.64         \\
    WCS 51	 &  VCC 651	          &   12:23:34.53	&   +09:30:54.9	  &  -1  	&   1	  &  17.55	 &   17.02	&  0.53	  &  0.62	              &      30.6	&   1.06         \\
    WCS 52	 &  VCC 1021	      &   12:27:33.48	&   +10:00:13.5	  &  12  	&   0	  &  15.54	 &   14.81	&  0.73	  &  0.09	              &      69.6	&   1.38         \\
    WCS 53	 &  VCC 1013	      &   12:27:30.16	&   +09:20:28.9	  &  12  	&   0	  &  16.84	 &   16.25	&  0.59	  &  0.43	              &      43.8	&   1.10          \\
    WCS 55\tablefootmark{a}	 &  VCCA 031	      &   12:25:31.46	&   +11:09:29.8	  &  12 	&   0	  &  17.50	 &   17.03	&  0.47	  &  1.96	              &      16.4	&   0.28         \\
    WCS 56\tablefootmark{a}	 &  VCCA 055          &   12:14:53.58	&   +09:40:14.4	  &  12	&   0	  &  17.25	 &   16.89	&  0.36	  &  0.80	              &      14.1	&   0.62             \\
    \hline                  
    \end{tabular}
    }
    \tablefoot{
    \tablefoottext{a}{Not a VCC member. Its optical diameter and morphological type were assigned manually.}
    \tablefoottext{b}{Galaxy with manually measured photometry.}
    }
\end{table*}    

\pagebreak
\section{Optical properties of \HI{} non-detected galaxies}
\begingroup  

\begin{longtable}{l c c c c c c c c}
    
    \caption{Catalog of spectroscopically confirmed galaxies detected optically but not in \HI{} in WAVES South. Coordinates show the optical position in J2000. The horizontal line divides the sample into two parts: first includes objects from the GOLDMine catalog and the second part includes objects from the NED query. All non-VCC galaxies in the second part had their morphological type assigned manually, as described in Sect.~\ref{section_optical_counterpart}.}   
    \small
    \label{nonHI_properties_table} \\    
    \hline\hline       
    Name                          &   R.A.		      &   Dec.	    	  &    Vel.	  &  Dist.	  &    Type	 &  $m_g$	 &   $m_i$	 &  $g-i$     \\
                                  &                   &                   & [$\rm{km\,s^{-1}}$]    & [Mpc]     &          &           &           &            \\
    \hline
    \endfirsthead
    \caption{Catalog of galaxies not detected in \HI{} in the WAVES South region (continued).} \\
    \hline\hline
    Name                          &   R.A.		      &   Dec.	    	  &    Vel.	  &  Dist.	  &    Type	 &  $m_g$	 &   $m_i$	 &  $g-i$     \\
                                  &                   &                   & [$\rm{km\,s^{-1}}$]    & [Mpc]     &          &           &           &            \\
    \hline
    \endhead
    \hline
    \endfoot
    
    VCC 118	                          &   12:14:36.80	  &   +09:41:22.0	  &    1292	  &  17.0	  &    -1	 &   16.15	 &   15.40	 &   0.75     \\
    VCC 216	                          &   12:17:01.10	  &   +09:24:27.0	  &    1325	  &  32.0	  &    -1	 &   14.99	 &   14.08	 &   0.91     \\
    VCC 227	                          &   12:17:14.50	  &   +08:56:32.0	  &    1304	  &  32.0	  &    10	 &   15.29	 &   14.48	 &   0.81     \\
    VCC 275	                          &   12:18:11.00	  &   +09:29:59.0	  &    1733	  &  32.0	  &    12	 &   14.77	 &   14.15	 &   0.63     \\
    VCC 394	                          &   12:20:08.60	  &   +09:28:05.0	  &    1789	  &  32.0	  &    -1	 &   17.38	 &   16.67	 &   0.71     \\
    VCC 407	                          &   12:20:18.80	  &   +09:32:43.0	  &    2078	  &  17.0	  &    -2	 &   14.54	 &   13.64	 &   0.89     \\
    VCC 458	                          &   12:21:12.70	  &   +08:57:47.0	  &    1563	  &  23.0	  &    -1	 &   16.85	 &   15.98	 &   0.86     \\
    VCC 504	                          &   12:21:50.10	  &   +09:44:21.0	  &    497	  &  17.0	  &    -1	 &   16.90	 &   16.21	 &   0.69     \\
    VCC 529	                          &   12:22:08.60	  &   +09:53:40.0	  &    1563	  &  17.0	  &    -1	 &   17.60	 &   16.77	 &   0.82     \\
    VCC 546	                          &   12:22:21.60	  &   +10:36:07.0	  &    2067	  &  17.0	  &    -1	 &   15.73	 &   14.97	 &   0.76     \\
    VCC 695	                          &   12:24:05.30	  &   +10:04:04.0	  &    1370	  &  23.0	  &    -1	 &   15.97	 &   15.15	 &   0.82     \\
    VCC 698	                          &   12:24:05.00	  &   +11:13:05.0	  &    2106	  &  17.0	  &    1	 &   12.97	 &   11.88	 &   1.09     \\
    VCC 747	                          &   12:24:47.80	  &   +08:59:29.0	  &    955	  &  23.0	  &    -1	 &   17.41	 &   16.55	 &   0.86     \\
    VCC 756	                          &   12:24:53.10	  &   +09:29:37.0	  &    728	  &  23.0	  &    -1	 &   17.09	 &   16.42	 &   0.67     \\
    VCC 856	                          &   12:25:57.90	  &   +10:03:14.0	  &    972	  &  23.0	  &    -1	 &   14.29	 &   13.35	 &   0.94     \\
    VCC 920	                          &   12:26:34.80	  &   +09:58:54.0	  &    443	  &  23.0	  &    -1	 &   17.44	 &   16.81	 &   0.63     \\
    VCC 931	                          &   12:26:44.00	  &   +10:54:17.0	  &    559	  &  17.0	  &    -1	 &   16.37	 &   15.54	 &   0.83     \\
    VCC 944	                          &   12:26:50.60	  &   +09:35:03.0	  &    832	  &  23.0	  &    1	 &   11.74	 &   10.58	 &   1.17     \\
    VCC 949	                          &   12:26:54.50	  &   +10:39:57.0	  &    1276	  &  23.0	  &    -1	 &   15.49	 &   14.50	 &   0.99     \\
    VCC 1003	                      &   12:27:26.50	  &   +11:06:27.0	  &    1130	  &  17.0	  &    2	 &   11.38	 &   10.06	 &   1.32     \\
    VCC 1039	                      &   12:27:44.30	  &   +11:12:52.0	  &    1442	  &  17.0	  &    -1	 &   17.14	 &   16.20	 &   0.94     \\
    VCC 1062	                      &   12:28:03.90	  &   +09:48:13.0	  &    517	  &  23.0	  &    1	 &   11.12	 &   9.91	 &   1.21     \\
    VCC 1075	                      &   12:28:12.30	  &   +10:17:52.0	  &    1844	  &  23.0	  &    -1	 &   14.72	 &   13.84	 &   0.88     \\
    VCC 1076	                      &   12:28:12.80	  &   +10:31:34.0	  &    1828	  &  17.0	  &    -1	 &   16.94	 &   16.08	 &   0.86     \\
    VCC 1078	                      &   12:28:11.30	  &   +09:45:37.0	  &    475	  &  23.0	  &    -1	 &   15.75	 &   15.33	 &   0.42     \\
    VCC 1079	                      &   12:28:12.00	  &   +10:21:55.0	  &    1468	  &  23.0	  &    -1	 &   16.97	 &   16.12	 &   0.85     \\
    VCC 1164	                      &   12:29:08.10	  &   +09:26:38.0	  &    1040	  &  23.0	  &    -1	 &   16.69	 &   15.85	 &   0.84     \\
    VCC 1209	                      &   12:29:40.60	  &   +10:23:06.0	  &    1454	  &  17.0	  &    -1	 &   17.29	 &   16.46	 &   0.84     \\
    VCC 1261	                      &   12:30:10.30	  &   +10:46:46.0	  &    1850	  &  17.0	  &    -1	 &   13.30	 &   12.39	 &   0.92     \\
    VCC 1273	                      &   12:30:16.90	  &   +09:05:06.0	  &    2015	  &  23.0	  &    12	 &   15.06	 &   14.36	 &   0.70      \\
    VCC 1303	                      &   12:30:40.60	  &   +09:00:56.0	  &    875	  &  23.0	  &    1	 &   12.80	 &   11.65	 &   1.15     \\
    VCC 1377	                      &   12:31:39.00	  &   +10:50:26.0	  &    634	  &  17.0	  &    12	 &   16.44	 &   15.67	 &   0.77     \\
    VCC 1412	                      &   12:32:06.20	  &   +11:10:35.0	  &    1342	  &  17.0	  &    3	 &   11.84	 &   10.64	 &   1.19     \\
    VCC 1422	                      &   12:32:14.20	  &   +10:15:05.0	  &    1372	  &  17.0	  &    0	 &   13.43	 &   12.52	 &   0.92     \\
    VCC 1444	                      &   12:32:35.90	  &   +09:53:11.0	  &    1769	  &  17.0	  &    -1	 &   15.96	 &   15.09	 &   0.87     \\
    VCC 1446	                      &   12:32:39.00	  &   +10:05:31.0	  &    2235	  &  17.0	  &    -1	 &   15.98	 &   15.07	 &   0.91     \\
    VCC 1481	                      &   12:33:09.00	  &   +10:50:10.0	  &    1957	  &  17.0	  &    -1	 &   17.76	 &   16.97	 &   0.79     \\
    VCC 1488	                      &   12:33:13.40	  &   +09:23:50.0	  &    1157	  &  17.0	  &    0	 &   14.84	 &   14.22	 &   0.62     \\
    VCC 1496	                      &   12:33:18.80	  &   +09:07:14.0	  &    1303	  &  17.0	  &    -1	 &   17.89	 &   17.04	 &   0.85     \\
    VCC 1509	                      &   12:33:31.60	  &   +09:27:33.0	  &    817	  &  17.0	  &    -1	 &   16.76	 &   15.91	 &   0.85     \\
    VCC 1521	                      &   12:33:45.00	  &   +10:59:45.0	  &    1212	  &  17.0	  &    0	 &   13.69	 &   12.59	 &   1.10      \\
    VCC 1549	                      &   12:34:14.80	  &   +11:04:18.0	  &    1357	  &  17.0	  &    -1	 &   14.46	 &   13.41	 &   1.05     \\
    VCC 1567	                      &   12:34:31.30	  &   +09:37:24.0	  &    1440	  &  17.0	  &    -2	 &   14.58	 &   13.58	 &   1.00      \\
    VCC 1629\tablefootmark{c}          &   12:35:38.00	  &   +09:35:29.0	  &    790	  &  17.0	  &    -1	 &   17.27	 &   16.43	 &   0.83     \\
    VCC 1661	                      &   12:36:24.80	  &   +10:23:05.0	  &    1400	  &  17.0	  &    -1	 &   16.14	 &   15.13	 &   1.01     \\
    VCC 1684	                      &   12:36:39.40	  &   +11:06:07.0	  &    694	  &  17.0	  &    -3	 &   15.08	 &   14.42	 &   0.66     \\
    VCC 1720	                      &   12:37:30.50	  &   +09:33:18.0	  &    2284	  &  17.0	  &    1	 &   12.17	 &   10.99	 &   1.18     \\
    VCC 1743	                      &   12:38:06.80	  &   +10:04:56.0	  &    1279	  &  17.0	  &    -1	 &   15.49	 &   14.61	 &   0.88     \\
    VCC 1767	                      &   12:38:37.60	  &   +09:40:31.0	  &    1472	  &  17.0	  &    -1	 &   16.64	 &   15.83	 &   0.82     \\
    VCC 1803	                      &   12:39:37.70	  &   +10:58:33.0	  &    1336	  &  17.0	  &    -1	 &   16.19	 &   15.28	 &   0.90      \\
    VCC 1813	                      &   12:39:55.90	  &   +10:10:34.0	  &    1834	  &  17.0	  &    3	 &   11.50	 &   10.17	 &   1.32     \\
    VCC 1826	                      &   12:40:11.20	  &   +09:53:46.0	  &    2033	  &  17.0	  &    -1	 &   15.56	 &   14.64	 &   0.92     \\
    VCC 1857	                      &   12:40:53.10	  &   +10:28:34.0	  &    634	  &  17.0	  &    -1	 &   14.99	 &   14.24	 &   0.75     \\
    VCC 1861	                      &   12:40:58.50	  &   +11:11:04.0	  &    683	  &  17.0	  &    -1	 &   14.50	 &   13.48	 &   1.02     \\
    VCC 1869	                      &   12:41:13.30	  &   +10:09:21.0	  &    1864	  &  17.0	  &    2	 &   11.88	 &   10.67	 &   1.22     \\
    VCC 1891	                      &   12:41:48.90	  &   +11:11:29.0	  &    1016	  &  17.0	  &    -1	 &   16.96	 &   16.03	 &   0.92     \\
    VCC 1895	                      &   12:41:52.00	  &   +09:24:10.0	  &    1032	  &  17.0	  &    -1	 &   14.93	 &   14.01	 &   0.92     \\
    VCC 1896	                      &   12:41:54.60	  &   +09:35:05.0	  &    1731	  &  17.0	  &    -3	 &   14.93	 &   13.93	 &   1.00      \\
    VCC 1919	                      &   12:42:18.90	  &   +10:34:04.0	  &    1869	  &  17.0	  &    -1	 &   16.99	 &   16.18	 &   0.81     \\
    VCC 1936	                      &   12:42:46.00	  &   +09:30:38.0	  &    985	  &  17.0	  &    -3	 &   15.70	 &   14.78	 &   0.92     \\
    VCC 1948	                      &   12:42:58.00	  &   +10:40:55.0	  &    1672	  &  17.0	  &    -1	 &   15.50	 &   14.65	 &   0.85     \\
    VCC 1958	                      &   12:43:10.30	  &   +11:02:12.0	  &    1049	  &  17.0	  &    -1	 &   16.71	 &   15.83	 &   0.88     \\
    VCC 1971	                      &   12:43:30.90	  &   +11:02:50.0	  &    1376	  &  17.0	  &    -1	 &   16.47	 &   15.57	 &   0.90      \\
    VCC 2000	                      &   12:44:32.00	  &   +11:11:26.0	  &    1115	  &  17.0	  &    0	 &   11.63	 &   10.47	 &   1.16     \\
    VCC 2042	                      &   12:46:38.40	  &   +09:18:25.0	  &    1765	  &  17.0	  &    -1	 &   15.19	 &   14.42	 &   0.76     \\
    VCC 2048	                      &   12:47:15.30	  &   +10:12:12.0	  &    1095	  &  17.0	  &    -3	 &   13.56	 &   12.58	 &   0.98     \\
    \noalign{\smallskip} \hline \noalign{\smallskip}                                                                                                  
    VCC 1357\tablefootmark{a,c}         &   12:31:24.42	  &   +09:28:28.0	  &    607	  &  23.0	  &    12	 &   18.93	 &   18.32	 &   0.61     \\
    VCC 1804\tablefootmark{a}          &   12:39:40.14	  &   +09:23:55.8	  &    1884	  &  17.0	  &    16	 &   15.81	 &   15.15	 &   0.66     \\
    VCC 1991\tablefootmark{b}          &   12:44:09.37	  &   +11:10:35.8	  &    1660	  &  17.0	  &    -1	 &   15.72	 &   14.70	 &   1.01     \\
    VCC 2012\tablefootmark{a}          &   12:45:05.66	  &   +10:54:03.2	  &    1066	  &  17.0	  &    -1	 &   14.65	 &   14.05	 &   0.61     \\
    VCC 2045\tablefootmark{b}          &   12:46:55.48	  &   +10:10:56.7	  &    1245	  &  17.0	  &    -1	 &   15.77	 &   14.87	 &   0.90      \\
    VCCA 038	                      &   12:30:02.62	  &   +09:24:11.9	  &    899	  &  23.0	  &    0	 &   17.33	 &   16.55	 &   0.78     \\
    SDSS J124101.06+094306.3    	  &   12:41:01.07	  &   +09:43:06.3	  &    1361	  &  17.0	  &    12	 &   17.75	 &   17.32	 &   0.43     \\
    WISEA J121341.44+090040.4	      &   12:13:41.43	  &   +09:00:40.3	  &    1801	  &  32.0	  &    0	 &   17.54	 &   16.70	 &   0.84     \\
    WISEA J121443.02+103159.2	      &   12:14:43.04	  &   +10:31:59.1	  &    2086	  &  17.0	  &    12	 &   18.07	 &   17.59	 &   0.49     \\
    WISEA J122020.13+094750.1	      &   12:20:20.14	  &   +09:47:50.2	  &    1189	  &  23.0	  &    12	 &   17.78	 &   17.32	 &   0.46     \\
    WISEA J123955.45+095520.4	      &   12:39:55.46	  &   +09:55:20.7	  &    1434	  &  17.0	  &    12	 &   17.72	 &   17.33	 &   0.39     \\
    WISEA J124533.24+102925.8	      &   12:45:33.30	  &   +10:29:25.7	  &    986	  &  17.0	  &    0	 &   18.29	 &   17.44	 &   0.85     \\
    \hline
\end{longtable}
\tablefoot{
\tablefoottext{a}{Present in the GOLDMine catalog, initially discarded due to low quality flags (larger than 3). Later added by the NED query but the morphological type is taken from GOLDMine.}
\tablefoottext{b}{Present in the GOLDMine catalog, initially discarded due to absent velocity measurements. Later added by the NED query but the morphological type is taken from GOLDMine}
\tablefoottext{c}{Manually measured photometry using SDSS \texttt{fits} files.}
}
\endgroup

\end{appendix}

\end{document}